\documentclass[pss]{wiley2sp} 
\usepackage{amsmath}
\usepackage{amsfonts}
\usepackage{amssymb}

\begin{document}

\title{A spin glass perspective on ferroic glasses.}
\titlerunning{Pseudo-spin glasses }

\author{%
  David Sherrington
}

\authorrunning{David Sherrington}

\mail{e-mail
  \textsf{D.Sherrington1@physics.ox.ac.uk}, Phone:
  +44-1865-273693, Fax: +44-1865-273947}

\institute{%
Rudolf Peierls Centre for Theoretical Physics, 1 Keble Rd., Oxford OX2 9EY, UK,\\
New College,  Oxford OX1 3BN, UK, \\
Santa Fe Institute, 1399 Hyde Park Rd., Santa Fe, NM 87501, USA.
}
\received{XXXX, revised XXXX, accepted XXXX} 
\published{XXXX} 

\keywords{Spin glass, relaxor, strain glass, disorder, frustration, complex system.}

\abstract{%
%
%
%

\abstcol{%
%
\vspace{7 mm}
A range of ferroic glasses, magnetic, polar, relaxor and strain glasses, are considered together from the perspective of spin glasses. Simple mathematical modelling  is shown to provide a possible conceptual unification to back similarities of experimental observations, without considering all possible complexities and alternatives.
  }{%
}}

%
%

\titlefigure[height=3.9cm, width=0.99\linewidth]{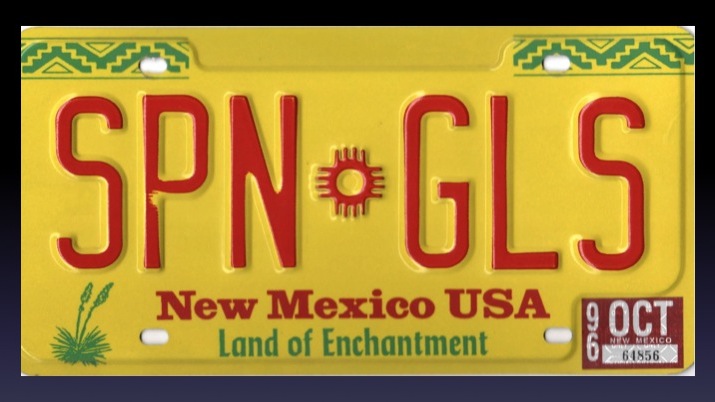}
\titlefigurecaption{%
}

%
%

\maketitle   

\section{Introduction}
\subsection{Spin glasses I}

The term `spin glass' was coined at the end of the 1960s to describe a set of alloys of mixed magnetic and non-magnetic metals, discovered earlier in the decade, that exhibited apparent cooperative  but non-periodic quasi-freezing of magnetic moments beneath a characteristic temperature  where the magnetic  susceptibility showed a peak; hence `spin' for magnetic moment and `glass' for non-periodic `freezing' and sluggishness. 

%
\begin{figure}[t]%
\center
\includegraphics*[width=\linewidth,height=0.53\linewidth]{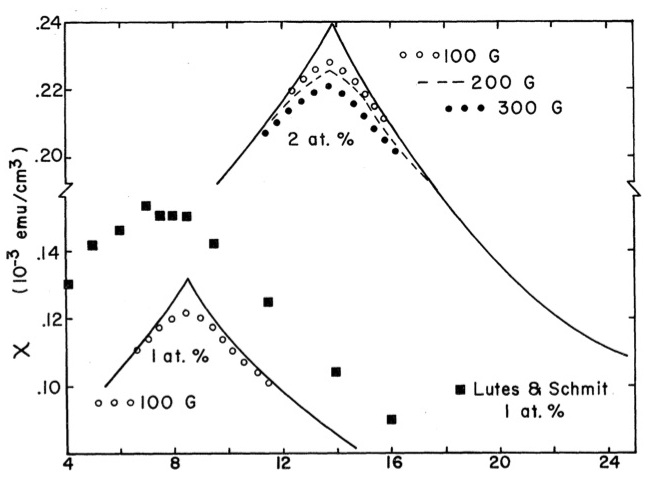}
\caption{%
Susceptiblity of dilute AuFe alloys for zero and several applied fields;
reprinted  with permission from V.~Cannella and
 J.\,A.~Mydosh,
 Phys.Rev.B \textbf{6}, 4220 (1971). \copyright 1971
American Physical Society. http://link.aps.org/abstract/PRB/v6/p4220} 
%
\label{fig:Cannella}
\end{figure} 
\begin{figure*}[htb]%
\center
\includegraphics*[width=\linewidth,height=0.35\linewidth]{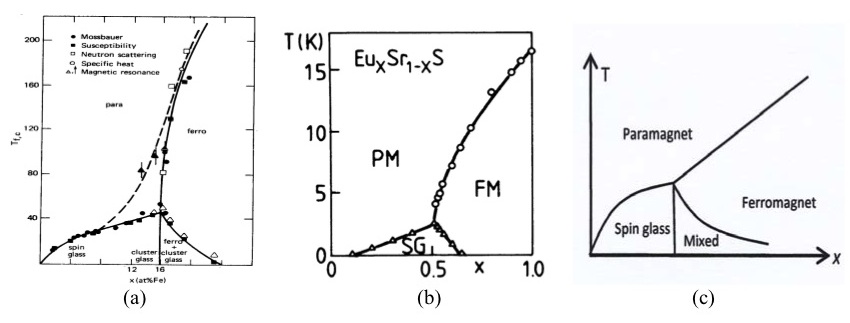}
\caption{%
Spin glass phase diagrams; (a) metal: AuFe; reprinted with permission from 
B.\,R.~Coles, B.~Sarkissian and R.\,H.~Taylor, Phil.Mag.B {\textbf{37}}, 489 (1978)
\copyright 1978 {Taylor and Francis Ltd. http://www.tandf.co.uk/journals},
(b) semi-conductor: EuSrS; reprinted with permission from  
H.~Maletta and P.~Convert, Phys.Rev.Lett. \textbf{42}, 108 (1979)
\copyright  1979 {American Physical Society}
http://link.aps.org/abstract/PRL/v42/p108,
(c) soluble model: SK Ising model with mean and variance of exchange distribution scaling with concentration $x$ \cite{foot-scaling}; from \cite{Sherrington_Glassy} \copyright {Springer-Verlag} (http://www.springer.com/). \label{fig:Sherrington_Glassy} }
\label{fig:phase_diagrams}
\end{figure*}

Initially these systems were viewed as `just' acquiring  slowed-down dynamics when the peak temperature was reached,  rather than exhibiting a true phase transition. However, early in the 1970s Cannella and Mydosh \cite{Cannella} demonstrated that if these systems were carefully isolated from external magnetic fields and only a small probe field is used then the susceptibility peak becomes sharp, a signal of a possible true phase transition; see Fig. {\ref{fig:Cannella}. This excited some theorists, leading Edwards and Anderson \cite{EA} to suggest a revolutionary new way to model and study such a transition in a highly original type of mean field theory. This in turn led to  a model for which a mean field solution is exact \cite{SK} but whose full solution has turned out to be extremely challenging and very subtle. Attempts to solve it have in turn led to major new conceptualization, new mathematical techniques and a plethora of extensions and applications \cite{DS-Ancona}, \cite{DS_RS}, \cite{MPV}, \cite{Stein-Newman}, many far in na\"{i}ve appearance from the original systems.

On the other hand, while the conceptual and mathematical attempts to understand the fundamental physics of spin glasses have had profound and far reaching effects, the actual materials that spawned this explosion of activity have not  themselves had practical application; for an experimental review more centred on the original magnetic systems see \cite{Mydosh}; for other classic reviews centred around these systems but also with theoretical modelling see \cite{Binder-Young} \cite{Fischer-Hertz}. A selection of history and development can be found in \cite{Stealing}.

In this article, arguments will be presented that  other classes of materials systems, 
including one that was discovered before the spin glass systems and that has received significant practical application and another that is a more recently considered variant of a different long-known  set of systems that have also received significant application, are in fact effectively quasi spin glasses, with corresponding consequences for understanding and for future examination. Again mathematical modelling will underlie the conceptual transfers, but in this case particularly within the framework of experimental spin glasses, rather than the statistical-physics-based extensions of the usual theorists' spin glass models that have been used in hard optimization, information theory and econophysics \cite{MPV} \cite{Mezard-Montanari} \cite{Coolen-neural} \cite{Coolen-MG}, and which have also stimulated mathematicians \cite{Talagrand} and mathematical physicists \cite{Bovier}.

\begin{figure*}[htb]%
\hspace{0.2cm}
\subfloat[ 
CuMn; FC  
((a) and (c)), 
ZFC
((b) and (d)); reprinted   with permission from 
S.~Nagata, P.\,H.~Keesom and H\,.R.~Harrison, Phys.Rev. B \textbf{19}, 1633 (1979)
\copyright 1979
American Physical Society. http://link.aps.org/abstract/PRB/v19/p1633]
{%
\includegraphics*[width=.47\textwidth,height=.305\textwidth]{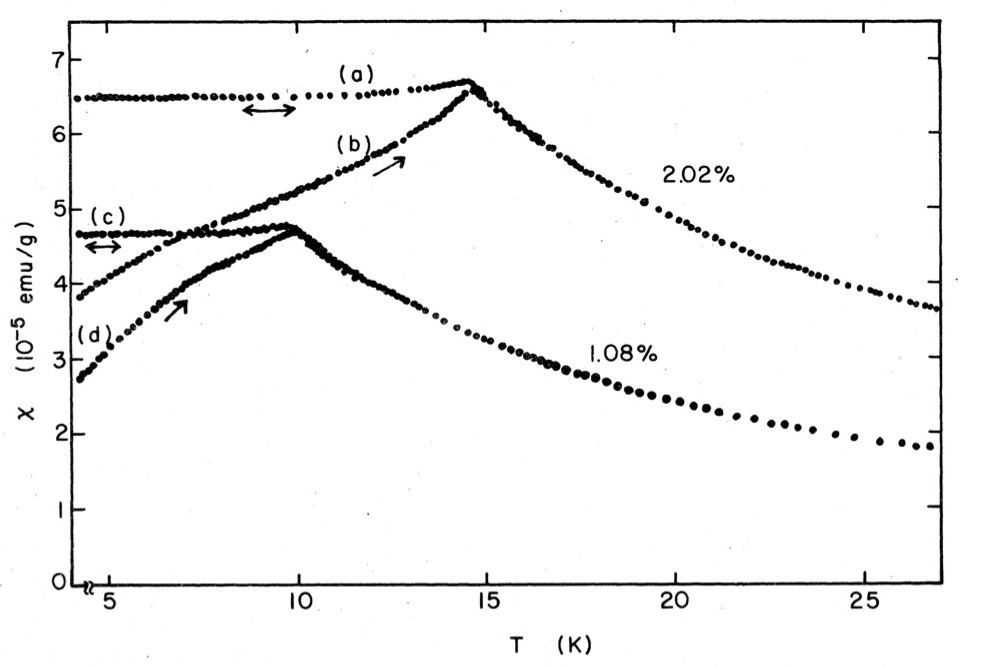}}\hfill
\subfloat[
BZT(50:50)  
from first-principles calculation/simulation:  
$\blacktriangle$:direct; 
$\circ$: ${\mathrm{fluctuation-dissipation}}$; 
reprinted  with permission from
A.\,R.~Akbarzadeh, S.~Prosandeev, E.\,J. ~Walter, A.~Al-Barakaty and L.~Bellaiche, {Phys.Rev.Lett.} {\textbf {108}}, 257601 (2012) 
 \copyright {2012 American Physical Society} 
http://link.aps.org/abstract/PRL/v108/p257601
]
{%
\includegraphics*[width=.490\textwidth,height=.305\textwidth]{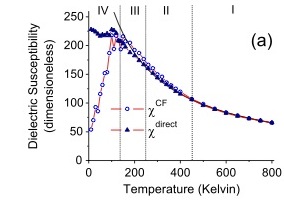}}\hspace{2.5cm}
\caption{%
FC and ZFC  susceptibilities in  spin glass and  relaxor.}
\label{fig:FC/ZFC susc}
\end{figure*}

\subsection{Spin glasses II}
The canonical spin glass systems, such as ${\textbf{Au}}$Fe and ${\textbf{Cu}}$Mn, are substitutional alloys whose magnetic  Hamiltonians may be expressed in the form 
\begin{equation}
H_{CSG}=-\sum_{(ij)(Mag)}  {J}({\textit{\textbf{R}}}_{ij}){\textit{\textbf{S}}}_{i}.{\textit{\textbf{S}}}_{j}
\label{eq:Hcsg}
\end{equation}
where the ${\textbf{S}}_{i}$ are localised spins on the magnetic ions (in the examples above, Fe and Mn), 
${J}({\textit{\textbf{R}}})$ is a translationally-invariant but
spatially-frustrated `exchange interaction' and the sum is over pairs of different sites, restricted to those occupied by magnetic atoms. Since these systems are metals, the exchange interaction has the RKKY form 
\begin{equation}
{J}({\textit{\textbf{R}}}_{ij}) ={\mathcal J}^{2}{\chi_{ij}}
\label{eq:RKKY}
\end{equation}
where $\mathcal{J}$ is the coupling strength between the conduction electron spin (${\textit{\textbf{s}}}_{i}$) and the local moment spin (${\textit{\textbf{S}}}_{i}$) ($H_{coupling} = {\mathcal{J }} {\textit{\textbf{s}}}_{i}.{\textit{\textbf{S}}}_{i}$) and $\chi_{ij}$ is the conduction band susceptibility as a function of separation between sites $i$ and $j$. Thus ${J}({\textit{\textbf{R}})}$ has a form that oscillates in sign with separation, with wavevector $2k_F$, where $k_F$ is the Fermi wavevector, and decays as $R^{-3}$. 

Edwards and Anderson \cite{EA} identified the key ingredients of  Eq.($\ref{eq:Hcsg}$) as the combination of competing interactions, now referred to as `frustration' \cite{Toulouse}, and quenched disorder. For these canonical experimental systems the frustration is due to the spatial oscillation  in sign of ${J}({\textbf{R}})$ \cite{footnote1}  and the randomness is that of the site occupation by magnetic or non-magnetic atoms. With this recognition they considered instead a system  involving only magnetic atoms but with the combination of spatial frustration and site disorder  of Eq.($\ref{eq:Hcsg}$) replaced by an effective system with a spin on every  site but with frustrated  bond disorder:  
\begin{equation}
H_{EA} = -\sum_{(lm)} J_{lm}{\textit{\textbf{S}}}_{l}{\textit{\textbf{S}}}_{m} 
\label{H_EA},
\end{equation}
with the  $J_{lm}$ chosen randomly from a distribution of mean $J_{0}$ and variance $J^{2}$. For $J_{0}/J$ large enough, the low temperature state is a ferromagnet; for the opposite case, beneath a critical value, the low temperature state is spin glass; while for higher temperature the phase is paramagnetic. With a further assumption that the interaction distribution is the same for all pairs of sites, independent of their separation \cite{SK}, this problem (known as the Sherrington-Kirkpatrick (SK) model) becomes soluble, but its solution is very subtle \cite{MPV} (as well as exciting)  and will not be a main consideration here \cite{foot_SK}.

Most theoretical and computational work has subsequently been performed on the random bond model, but in this paper we shall principally compare other material systems  with the random-site, spatially-frustrated experimental spin glass systems. 

An early clear recognition of the  potentiality for spin glass behaviour of other combinations of frustration and disorder came from the appreciation that spin glasses need not be metals, provided there is a different source of frustration. Such a situation was found in ${\rm{Eu}}_{x}{\rm{Sr}}_{1-x}{\rm{S}}$ \cite{Maletta}, a semiconductor where only the Eu is magnetic but there is competition between nearest ferromagnetic coupling and next nearest neighbour antiferromagnetic super-exchange. Fig. $\ref{fig:phase_diagrams}$ shows phase diagrams for (a) metallic spin glass, (b) semi-conducting spin glass, and (c) the soluble SK (Ising) spin glass with concentration scaling emulating site dilution \cite{foot-scaling}; the region labelled `Mixed' is a glassy ferromagnet.

Interesting experimental manifestations of conventional spin glasses include \newline (i) a peak in the static zero-field-cooled  susceptibility in the small-field limit  at a critical temperature $T_f$ \cite{Cannella} (Fig  {\ref{fig:Cannella}); \newline
 (ii)  non-ergodicity  beneath $T_f$, shown up  by 
\newline 
\indent
 {(a)  onset of differences between field-cooled (FC) and zero-field-cooled (ZFC) susceptibilities at $T_f$, FC plateauing beneath $T_f$ \cite{Nagata} %
(Fig \ref{fig:FC/ZFC susc}a),} 
 and \newline
\indent
(b) differences between magnetizations remaining at low temperatures, beneath $T_f$, after applying and then removing an applied magnetic field, the thermoremanent magnetization (TRM)
 being greater than the isothermal remanent magnetization (IRM) \cite{foot-IRM-TRM},
 merging at sufficiently high fields to a non-zero value \cite{Tholence} \cite{Nordblad}}
(Fig $\ref{fig:TRM_IRM}$); \newline
 (iii) temperature-shifts in the positions of peaks in the a.c. susceptibility, higher frequencies showing peaking at higher temperatures \cite{Wassermann} (Fig $\ref{fig:a.c.susc}$a);  \newline (iv)  aging, re-juvenation and memory
\cite{Vincent}\cite{Jonason} (Fig $\ref{fig:Vincent}$).

\begin{figure*}[htb]%
\hspace{1cm}
\subfloat[
 AuFe; reprinted with permission from  
J.\,R.~Tholence and R.~Tournier, J.Physique Colloq. \textbf{35}, C4-229 (1974)
\copyright 1974  {EDP Sciences} (http://edpsciences.org/) .] 
{%
\includegraphics*[width=.40\textwidth,height=.25\textwidth]{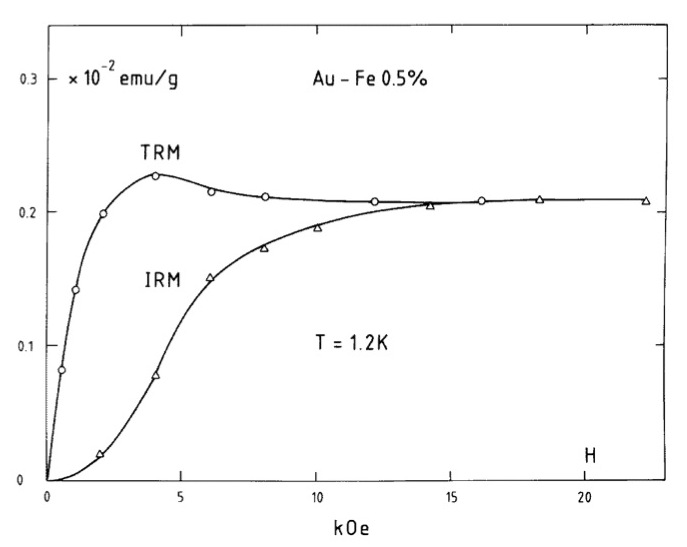}}
\hfill
\subfloat[CuMn; reprinted with permission from  
P.~Nordblad, L.~Lundgren and L.~Sandlund, Europhys.Lett. {\textbf{3}}, 235 (1987) 
\copyright  1987 {EDP Sciences} (http://edpsciences.org/) .] 
{%
\includegraphics*[width=.40\textwidth,height=.25\textwidth]{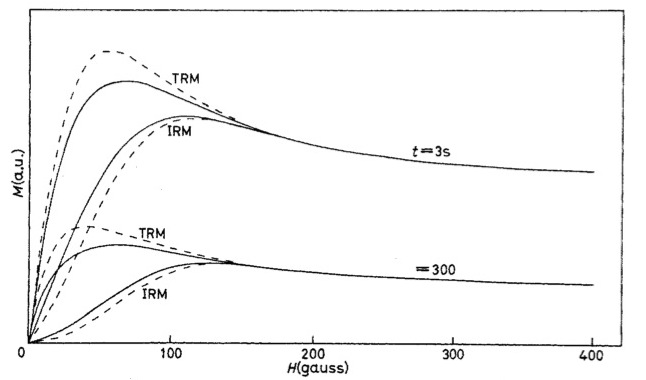}} \hspace{.5cm}
\caption
{%
Thermoremanent magnetization (TRM) and isothermal remanent magnetization (IRM) in spin glasses.}
\label{fig:TRM_IRM}
\end{figure*}

While the spins in the canonical spin glasses mentioned above are hard, of essentially fixed length (albeit with extra quantum effects and excitations that we have not discussed and consider secondary for our present purpose), analytic theoretical studies often replace them by the limits of soft spins, usually for the EA random-bond case  
\cite{deDominicis}. Such a soft-spin formalism allows extensions that will be relevant below, particularly clearly within the random-site description. Thus one can extend Eq. ($\ref{eq:Hcsg}$) to 
\begin{eqnarray}
H_{SSG}=\sum_{i} \left \{
\kappa_{i}
{|
\textit{\textbf{S}}
|}^2
+ {\lambda_{i}} {|\textit{\textbf{S}}_{i}|}^4
+{\mu_{i}} {|\textit{\textbf{S}}}_{i}|^6 \right \}  \nonumber \\
 -\sum_{(ij)}  {J}({\textit{\textbf{R}}}_{ij}){\textit{\textbf{S}}}_{i}.{\textit{\textbf{S}}}_{j}
\label{eq:Hsg}
\end{eqnarray}
where now the spins ${\textit{\textbf{S}}}$ are soft, of indeterminate {\it{a priori}} length, and the sums are over all sites, but the $\{\kappa_{i}\}, \{\lambda_{i}\}$ and $\{\mu_{i}\}$ depend upon the types of atoms occupying the sites \{$i$\}. The $\mu$ are always taken as positive, to bound.

In the case of hard spins, their $\kappa$ are large and negative, their $\lambda$ are large and positive but their ratio ($|\kappa|/\lambda$) is constant.  Non-magnetic sites have $\kappa$, $\lambda$ and $\mu$ all very positive, so that their ${\textit{\textbf{S}}}$ are always quenched to zero. In the language of structural phase transitions, hard spins exhibit order-disorder transitions, both for periodic ordering, as for example when all sites are occupied by magnetic atoms and the order is determined by the overall favoured compromise, 
and for spatially disordered systems with non-periodic phases, such as spin glasses.


If   $\kappa$, $\lambda$ and $\mu$ on any site are all positive and there are no interactions, the ground state has the corresponding $\{{\textit{\textbf{S}}}={\textbf{0}}\}$. Ordering becomes possible at low enough temperature only if the energetic gain from the interaction term of suitable displacements overcomes the local energetic cost, with the actual ordering determined by the greatest gain. For a pure system this is the analogue of a displacive ferroelastic  ordering in structural phase transitions. A pure magnetic example of such behaviour lies in itinerant magnetism, although it is not usually expressed as above. It is however straightforward to consider a simple example in this language and see how it extends to a spin glass as well as a ferromagnet \cite{DS-Mihill}. 

Consider a simple binary Hubbard model alloy of atomic types $A$ and $B$;
\begin{equation}
H_{HA}=\sum_{ij;s=\uparrow,\downarrow} t_{ij} a_{is}^\dagger a_{js} +\sum_{i;s=\uparrow,\downarrow}V_{i} a_{is}^{\dagger}a_{is} + \sum_{i}U_{i}\hat{n}_{i\uparrow}\hat{n}_{i\downarrow}
\label{eq:Hubbard}
\end{equation}
where the $a,a^{\dagger}$ are site-labelled d-electron annihilation and creation operators, $\hat{n}_{is}=a^{\dagger}_{is}a_{is}$, and in general the $t_{ij}$, $V_{i}$ and $U_{i}$ depend upon whether the atoms at sites {\it{i,j}} are of type $A$ or type $B$. For simplicity let us assume that in fact the $t_{ij}$ can be taken to be independent of the occupation. Next we transform to local electron charge and local electron magnetization variables
\begin{equation}
n_{i}=n_{i\uparrow} + n_{i\downarrow}; ~~{\textit{\textbf{S}}}_{i}=a_{is}^{\dagger}{\boldsymbol{\sigma}}_{s,s'}a_{is'}
\end{equation}
and furthermore  (i) assume that the local charge contribution can be treated in mean field theory and (ii) simplify by taking $(V_{i} + U_{i}n_{i}/2)= constant$, leaving an effective  d-electron Hamiltonian 
\begin{equation}
H=\sum_{ij,\sigma}t_{ij}a_{i\sigma}^\dagger a_{j\sigma} -{\frac{1}{4}}\sum_{i}U_{i}{\textit{\textbf{S}}}_{i}.{\textit{\textbf{S}}}_{i}.
\end{equation}
Note that the ${\textit{\textbf{S}}}$ are still quantum operators. 

\begin{figure*}[htb]%
\hspace{1cm}
\subfloat[
PtMn spin glass; reprinted with permission from
 G.\,V.~Lecomte, H.~von L{\"o}hneysen and E.\,F.~Wassermann, {Z.Phys. B} {\textbf{50}}, 239 (1983)
\copyright 1983{Springer-Verlag} (http://www.springer.com/).]{%
\includegraphics*[width=.400\textwidth,height=.325\textwidth]{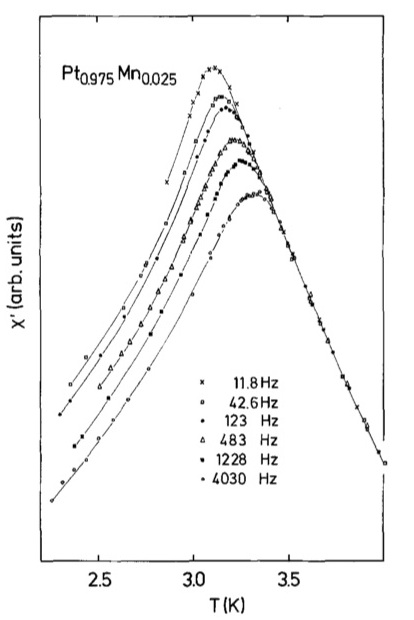}}\hfill
\subfloat[
$\mathrm {Ba(Zr}_{1-x}\mathrm{Ti}_{x}\mathrm{)O}_3$ relaxor 
($x=0.65$);  from  
T.~Maiti, R.~Guo and A.\,S.~Bhalla, {J.Am.Cer.Soc.} {\textbf 91}, 1769 (2008)
\copyright  {American Ceramics Society}
 ]{%
\includegraphics*[width=.480\textwidth,height=.325\textwidth]
{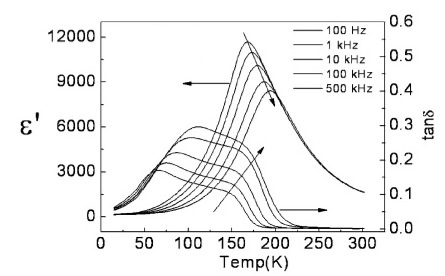}
}\hspace{1cm}%
\caption{%
Frequency-dependent a.c. susceptibilities of (a) a spin glass and (b) a relaxor.}
\label{fig:a.c.susc}
\end{figure*}

%
%

%
%

Writing the partition function as a functional integral, utilising the now-quadratic form of the U term to introduce local magnetization variables via the inverse of completing the square \cite{Gelfand} \cite{Stratonovich} \cite{Hubbard}, integrating out the original electron operators \cite{Sherrington2}, and taking the static approximation, 
one obtains an effective Hamiltonian in local magnetisation variables; to fourth order,
\begin{eqnarray}
&H_{m}= \sum_{i}  (1-U_{i}\chi_{ii})   
|{\textit{\textbf{m}}}_{i}|^2 - 
 \sum_{ij;i\neq j}U_{i}^{1/2}U_{j}^{1/2}\chi_{ij}{\textit{\textbf{m}}}_{i}.{\textit{\textbf{m}}}_{j} \nonumber \\
 &- 
\displaystyle{\sum_{ijkl;\alpha\beta\gamma\delta}}(U_{i}U_{j}U_{k}U_{l})^{1/2}\Pi^{\alpha\beta\gamma\delta}_{ijkl}
m^{\alpha}_{i}m^{\beta}_{j}m^{\gamma}_{k}m^{\delta}_{l},
\label{H_SM}
\end{eqnarray}
where $\chi$ is the static band susceptibility function of the bare system with all $U=0$ and $\Pi$ is a corresponding  bare 4-point function.

Consider a system in which $U^{A}=0$ but $U^{B}\neq{0}$. Pure $A$ has only magnetic fluctuations but no macroscopic order, while pure $B$ can have cooperative ferromagnetic order if $(1-U^{B}{\sum_{ij}\chi_{ij})}<0$;{\it{ i.e.}} if the Stoner criterion is satisfied \cite{Stoner}. A single $B$ atom at site $i$ in an $A$ host can only have a mean-field local moment if the Anderson condition $(1-{U^{B}}\chi_{ii})<0$  is satisfied \cite{Anderson61}.  Since $\chi_{ij}$ oscillates in sign as a function of separation, a more concentrated alloy, with a finite non-zero density of $B$ atoms, can exhibit either ferromagnetism, if the concentration of $B$ atoms is high enough, or spin-glass order, beneath a critical concentration $x_c$. If the Anderson local moment  criterion is  satisfied then the situation is essentially the same as in the conventional hard spin case discussed above. However, if the Anderson criterion is not satisfied then a sufficiently strong energy lowering due to coherently-acting spontaneous local magnetization fluctuations is needed to bootstrap a magnetically ordered phase, overcoming the $(1-U^{B}\chi_{ii}) |{\bf{m}}_{i}|^2$ local fluctuation penalty.  
For the case of a high concentation of $B$ this phase is still essentially Stoner's itinerant ferromagnetism. But for an intermediate concentration of $B$ the spontaneous cooperative phase can be a spin glass, bounded by a lower critical concentration separating it from the (Pauli-type) paramagnet and an upper critical concentration separating it from the ferromagnet.
 This sequence was already observed in the early days of experimental spin glass physics; {\it{e.g.}} in RhCo alloys \cite{Coles-Tari}.

\begin{figure*}[htb]%
\subfloat[
Out of phase susceptibility of ${\mathrm{CdCr}}_{1.7}{\mathrm{In}}_{0.3}{\mathrm{S}}_{4}$ 
spin glass during negative temperature cycling; exhibiting aging, rejuvenation and memory; 
reprinted with permission from
F.~Lefloch, J.~Hammann, M.~Ocio and E.~Vincent, Europhys.Lett. {\textbf{18}}, 647 (1992)
%
%
\copyright 1992  {EDP Sciences} (http://edpsciences.org/)).]{%
\includegraphics*[width=.330\textwidth,height=.25\textwidth]{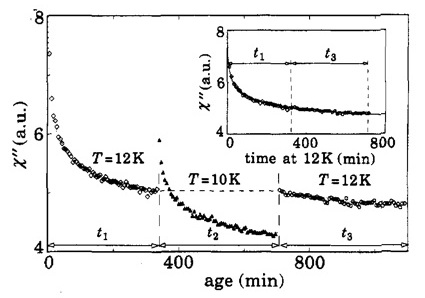}}\hfill
\subfloat[
Double-dip temperature cooling and reheating protocol used for Fig $\ref{fig:Vincent}$c; 
\copyright 2000 { EDP Sciences} (http://edpsciences.org/)   ]{%
\includegraphics*[width=.200\textwidth,height=.235\textwidth]{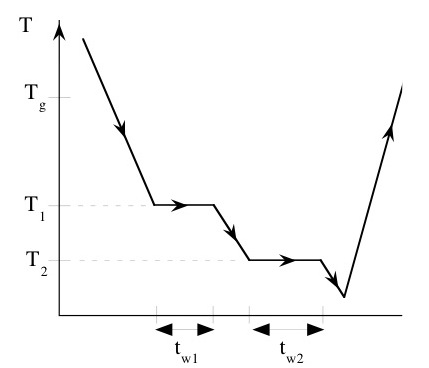}}
\hfill%
\subfloat[
Out-of-phase susceptibility vs. temperature following the protocol of Fig $\ref{fig:Vincent}$b;  
reprinted with permission  from 
K. ~Jonason, P.~ Nordblad, E. ~Vincent, J. ~Hammann and J.-P.~ Bouchaud,
Eur. Phys. J. B {\textbf{13}}, 99 (2000)
\copyright 2000 {EDP Sciences} (http://edpsciences.org/)]{%
\includegraphics*[width=.370\textwidth,height=.245\textwidth]{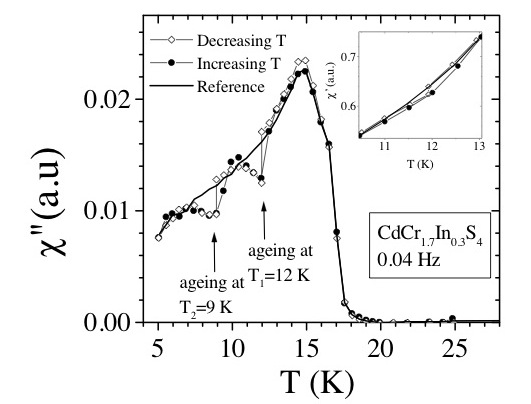}}%
\caption{%
Aging, re-juvenation and memory  in spin glasses.
}
\label{fig:Vincent}
\end{figure*}

In cases where both $U^A$ and $U^B$ are non-zero it is convenient to rescale and employ $M_{i}=U_{i}m_{i}$ as variables, so that all the disorder lies in the local harmonic term
\begin{eqnarray}
H_{M}= \sum_{i}  (U_{I}^{-1}-\chi_{ii})   
|{\textit{\textbf{M}}}_{i}|^2 - 
 \sum_{ij;i\neq j}\chi_{ij}{\textit{\textbf{M}}}_{i}.{\textit{\textbf{M}}}_{j} \nonumber \\
 - 
 \sum_{ijkl;\alpha\beta\gamma\delta}\Pi^{\alpha\beta\gamma\delta}_{ijkl}
M^{\alpha}_{i}M^{\beta}_{j}M^{\gamma}_{k}M^{\delta}_{l} + ...
\label{H_SM2}
\end{eqnarray}
This form will be useful in a comparison below.

In the lower paramagnetic concentration region of the $AB$ alloy there will still be statistical fluctuations in the occupation of atomic sites by $A$ or $B$ atoms and these will lead to the formation of cluster local moments, even in the paramagnetic phase, as a further simple extension of this modelling shows \cite{DS-Mihill}. Further discussion will, however, be postponed until later in this paper, in the section on relaxors.

A third situation of interest within the general model of 
eq. ($\ref{eq:Hsg}$) 
is where $\kappa>0$, $\lambda<0$ and $\mu>0$. If $\lambda$ is sufficiently negative compared with $\kappa$ then even isolated atoms, without any need for interaction, can have local moments; with interactions they behave analogously to the hard-spin systems discussed above. The extra interest is in situations where the $\lambda$ are not sufficiently negative for local moments in the absence of interactions. With interactions they can however exhibit first-order phase transitions to spontaneous cooperative magnetism. Again, with frustration and disorder the result can include spin glass phases. An early discussion, within the context of a spin-1 
Ising model,
\begin{equation}
H_{GS}=\sum_{i}D {S_{i}^2} 
-\sum_{ij}J_{ij}S_{i}S_{j}; ~~S_{i}=0,\pm{1},
\label{eq:GS}
\end{equation}
 with quenched random \{$J_{ij}$\}, can be found in \cite{DS-Ghatak}. For positive $D$, locally favouring $S=0$, a sufficient interaction bootstrapping energy is needed to induce moments and a cooperative phase \cite{Sherrington-induced-moment}. Clearly, one can have analogues with site randomness and spatial frustration.


\section{Relaxors}\label{section:Relaxors} 
In the 1950s, following interest in pure displacive ferroelectrics of perovskite structure and make-up ${AB\rm{O}}_3$, where $A$ is an ion of charge +2, $B$ an ion of charge +4 and O is oxygen, of charge -2,  a new class of alloy materials was discovered in $\mathrm {Pb(Mg}_{1/3}\mathrm{Nb}_{2/3}\mathrm{)O}_3$ (PMN)
\cite{Smolenskii} with interesting 
 frequency-dependent peaks in its dielectric susceptibilities as a function of temperature \cite{Smolenskii2}, just below room temperature, but without any macroscopic polarization in the absence of applied fields.  It further turned out that there are also 
higher temperature manifestations of nano-scale polar domains \cite{Egami}, as well as non-ergodicity \cite{Kleemann}, \cite{Kleemann_FC_ZFC}, \cite{Levstik}  beneath a temperature comparable with that of the finite-frequency susceptibility peaks.

The frequency-dependent peaking  of PMN and other since-discovered related systems, now known collectively as `relaxors',  has proven to be of significant application value, as well as fundamental interest; for reviews of the physics see \cite{Cross}, \cite{Samara}, \cite{Cowley}. However, although pictures have been proposed,  there is no universally accepted understanding of the origin of their behavior. 

PMN remains the most famous relaxor, but it is complicated by the fact that Mg and Nb are not isovalent, having charges respectively +2 and +5 and
giving rise to perturbing extra charges compared with the pure template +4 of the classic displacive ferroelectrics, 
   and hence giving rise to random fields.  By contrast, the more recently recognised perovskite alloy
$\mathrm {Ba(Zr}_{1-x}\mathrm{Ti}_{x}\mathrm{)O}_3$ (BZT),
has
 Zr and Ti   isovalent, both of charge +4 , yet still exhibits the characteristic relaxor frequency-dependent susceptibility peaks \cite{Maiti}, \cite{Shvartsman}, \cite{Kleemann-Miga} for a range of relative (Zr:Ti) concentations. Fig \ref{fig:a.c.susc}b shows that this behaviour is very reminiscent of that of a spin glass.
Recent first-principles computer simulations  have also exhibited the complementary relaxor features of the onset of non-ergodicity at a characteristic temperature and nano-domains for a range of temperatures above the non-ergodicity onset \cite{Akbarzadeh2012}. Again, the behaviour is very reminiscent of a spin glass, as illustrated in Fig \ref{fig:FC/ZFC susc}.

\subsection{Isovalent relaxors}

The present author has recently argued that the relaxor behavior of BZT can be understood by a mapping and analogy with spin glasses \cite{DS2013}, without any need to posit random fields, which have been a popular belief of the origin of relaxor behavior in PMN \cite{Kleemann}.

Athough quantum effects underlie some of the effective interactions in ferroelectrics, as they do in magnets, they can, in practice, be modelled classically in terms of the ionic displacements. Furthermore,  although all four types of ion in BZT ( Ba, Zr, Ti and O) can, in principle, move, we note that the main 
motion is of the $B$-type ions. Hence we absorb the effects of the $A$-type Ba and the O into an effective  Hamiltonian in terms of displacements on the $B$-sites. Further noting that the $B$-sites are occupied randomly by Ti or Zr and that these two ions have different local restoring coefficients, ignoring  for simplicity both local anisotropy and coupling to overall strain fields  (since they are expected not to be important for relaxors, which we shall argue have no overall strain \cite{foot-strain}, and as is also observed in numerical studies of BZT with $x=0.5$ \cite{Akbarzadeh2012}) , and replacing site-to-site interactions by an annealed average, we are left with a Hamiltonian of the form
\begin{equation}
H_{RF}=\sum_{i} 
\{\kappa_{i}| \textit{\textbf{u}}_{i} |^2
+\lambda_{i}|\textit{\textbf{u}}|^4    
\}+ {\sum_{ij} H^{avg}_{int}({\textit{\textbf{u}}}_{i}, {\textit{\textbf{u}}}_{j},  {\textit{\textbf{R}}}_{ij})}
\label{eq:H2}
\end{equation}
where the sites $\{i\}$  are occupied randomly by Ti or Zr, with relative probabilities {\it{x}}:(1-{\it{x}}) and with corresponding  $\kappa^{\rm{Ti}}, \kappa^{\rm{Zr}}, \lambda^{\rm{Ti}}$ and $\lambda^{\rm{Zr}}$, and $H^{avg}_{int}({\bf{u}}_{i}, {\bf{u}}_{j},{\textit{\textbf{R}}}_{ij})$ is the averaged site-to-site interaction.
The zero-temperature phase structure is given by minimising $H$ with respect to the $\{\textit{\textbf{u}}_{i}\}$. 

Eq. ($\ref{eq:H2}$) is recognisable as essentially the same as Eq. ($\ref{eq:Hsg}$), with the displacements playing the roles of soft pseudo-spins  with positive $\kappa$ and $\lambda$.  Consequently the expectations of the corresponding spin case pass over to the present displacement problem, particulary as epitomised by the model of Eq. ($\ref{H_SM2}$) with two different values of $U$, neither of them sufficient to satisfy the Anderson mean-field local moment criterion in isolation in a host with $U=0$ but one sufficient to bootstrap periodic cooperative order when it alone is present.

The crucial difference between systems with Zr or Ti at a B site lies in the strength of their $\kappa$; these are weak for Ti, permitting the low-temperature ferroelectric distortion 
observed in $\rm{ BaTiO_3}$, whereas in $\rm{ BaZrO_3}$ the Zr harmonic restoring coefficient  is  stronger and prevents macroscopic global distortion even down to zero temperature. The fact that $\kappa ^{\rm{Zr}}$ is so large  implies that in the alloys BZT all the sites ${\{i\}}$ occupied by Zr atoms have \{$\textit{\textbf{u}}_{i}=\textbf{0}$\} and hence may be ignored in considering the spontaneous displacements. Hence one is left with the effective Hamiltonian 
\begin{eqnarray}
H_{BZT} 
 = 
\sum_{i(Ti)}   \{\kappa^{\rm{Ti}}   {|{\textit{\textbf{u}}}_{i}|}^2 
+   {\lambda^{\rm{Ti}}} 
|\textit{\textbf{u}}_{i}|^4
\}
 \nonumber
 \\
 + 
{\sum_{ij(Ti)} } H_{int}(   {\textit{\textbf{u}}}_{i},    {\textit{\textbf{u}}}_{j},    {\textit{\textbf{R}}}_{ij}).
\label{eq:Heff}
\end{eqnarray}
with sums now restricted to  B-sites occupied by Ti ions.

The fact that experimentally the low temperature state of $\rm{ BaTiO_3}$ is ferroelectric shows that the dominant interaction in $H_{int}$ is ferroelectric. However, there are  both ferroelectric and anti-ferroelectric contributions at different separations \cite{Zhong}. Again this is a situation analogous to that for many conventional spin glasses, albeit with a different detailed form of the interactions \cite{foot:AF}. 


With this recognition it becomes clear that one should expect a sequence of low temperature phases with the concentration $x$ of Ti in $\mathrm {Ba(Zr}_{1 -x}\mathrm{Ti}_{x})\mathrm{O}_3$; ferroelectric ${\rightarrow}$ {pseudospin glass} ${\rightarrow}$ paraelectric for $x$ in the ranges $1 \ge {x}_{g}(T) \ge x_{p}(T) \ge 0$. The pseudospin glass is, of course,  the state usually known as a relaxor. Correspondingly, there will be transitions as a function of reducing temperature, paraelectric $\rightarrow$ ferroelectric at $T_{c}(x)$ for $x>x_c$ 
and paraelectric $\rightarrow$ relaxor
at $T_{g}(x)$
for 
$x_{c} \ge x \ge x_{p}$  \cite{foot-strain2}.

Again by analogy with spin glasses, (i) within the relaxor concentration range one would expect peaks in the frequency-dependent susceptibility at temperatures above $T_{g}(x)$, increasing with the frequency, as has been observed in BZT \cite{Maglione}  \cite{Maiti} \cite{Shvartsman} \cite{Kleemann-Miga}, and (ii)  $T_g(x)$ should mark the onset of non-ergodicity and preparation-dependence, the Zero-Field-Cooled (ZFC) susceptibility peaking and the Field-Cooled (FC) susceptibility `freezing', as found in the simulations of \cite{Akbarzadeh2012}, the FC susceptibility essentially measuring a full Gibbs average over all pure states while the ZFC susceptibility essentially restricts to a single pure macrostate \cite{MPV} \cite{Parisi}. 

The susceptibility values at their maxima become much larger as the concentration $x$ is increased towards the critical concentration for the onset of ferroelectricity, far beyond linearity with the concentration of Ti \cite{Maiti}. This is also a standard result for spin glasses \cite{SK}.

Another feature of relaxors that has elicited much interest is the observation of `polar nanodomains' (PNRs) for a significant temperature region above that of the peaks in the susceptibilities, seen directly for example in neutron scattering studies of PMN \cite{Egami} and more recently BZT \cite{Maiti}, as well as in the computer simulations of \cite{Akbarzadeh2012}, and understood as consequences of statistical fluctuations in the site-distributions of the constituent elements \cite{Cross}. The formation of PNRs becomes clear within a simple combination of the soft pseudospin modelling discussed above with Anderson's work on localization in disordered electronic systems \cite{Anderson1958}.


To see this $H_{RF}$ and $H_{BZT}$  are re-interpreted as Ginzburg-Landau free energies with their parameters renormalized as a function of temperature. The effective `local nanodomains' are given by minimization with respect to the \{$\textit{\textbf{u}}_{i}$\}, 
yielding values given in simple mean field theory by the self-consistent solution of
\begin{equation}
{\kappa^{T}_{i}} {\textit{\textbf{u}}_{i}} + 2{\lambda^{T}_{i}}   {\textit{\textbf{u}}}_{i}
+ {\sum_{j} \partial {  H^{T}_{int}({\textit{\textbf{u}}}_{i}, {\textit{\bf{u}}}_{j},  {\textit{\textbf{R}}}_{ij})}/\partial{\textit{\textbf{u}}}_{i} }=0
 {|    \textit{\textbf{u}}_{i}    |}^2,
\label{eq:self-cons}
\end{equation}
with the  superscripts $T$ indicating  `temperature renormalized'. The most important conceptual feature is that the $\{\tilde\kappa\}$ increase with increasing temperature relative to the interaction term, so that at high enough temperature paraelectricity is favoured for all $x$. Eqn (${\ref{eq:self-cons}}$) is then closely similar to the corresponding minimization of the Hamiltonian $H_{M}$ of the transition metal spin glass alloy. Following  \cite{DS-Mihill}  the formation of local nanodomains is thus conceptually relatable to an  Anderson eigen-equation exhibiting localization in a system with local electron potential disorder \cite{Anderson1958}.

For conceptual simplicity, considering a scalar displacement analogue of Eq.($\ref{eq:self-cons}$), ignoring spatial anisotropy, taking $\tilde H_{int}$ to be quadratic, and re-ordering, there results the simplified self-consistency equation
\begin{equation}
\kappa^{T}_{i} u_{i}
+ {\sum_{j}  J^{T}_{ij}u_{j}=
- 2\lambda^{T}_{i}} {u_{i}}^3.
\label{self-cons_2}
\end{equation}
This can usefully be compared with the Anderson eigen-equation
\begin{equation}
\epsilon_{i} \psi_{i} 
+\sum_{j}  t_{ij} \psi_{j} =E\psi_{i}.
\label{Anderson_equiv}
\end{equation}
Hence we see that with suitable identifications of the $\{ \epsilon_{i } \}$ 
with the $\{ {\kappa}^{T}_{i} \}$ 
and the $\{ t_{ij} \}$ 
with the $\{  J^{T}_{ij} \}$, one can  relate  mean field `cluster moments' in the displacive system to corresponding negative energy (E) solutions of the eigenequation.

However, solutions to Eq. ($\ref{Anderson_equiv}$) with quenched $\kappa$-disorder can be either extended or localized; extended in the centre of the band of eigenstates, localized states at both extremities, with the extended and  localized regions separated by lower $E_{m_L}$ and upper  $E_{m_U}$ `mobility edges'. Only extended states can give rise to a true cooperative distortive phase, periodic or otherwise. Thus a condition for cooperative order is that the $E_{m_L}$  must be negative, although, even if  $E_{m_L}>0$, there can still be localised states/polar nanodomains if there is density of states of Eq. ($\ref{Anderson_equiv}$) at $E<0$. 

Note that  the density of states, and with it the mobility edges, is temperature-dependent through the renormalization of the $\tilde\lambda$ and $\tilde J$, shifting to lower $E$ with decreasing $T$.   The onset of mean-field `cluster moments', observable on finite timescales as nano-domains, is thus given by the onset of  solutions to Eq.($\ref{Anderson_equiv}$) with $E \leq 0$ , while the true thermodynamic transition, which requires an extended state,   occurs only  when the mobility edge $E_{m_L}(x,T)$ becomes zero. Except for the pure limits, for which all eigenstates are extended, the critical temperature for the nano-domains is  higher  than that for a true phase transition.

In the usual Anderson situation the $\{t_{ij}\}\ge{0}$  so that extended states are ferroelectric. However, in a frustrated case with $\{ J^{T}_{ij}\}$ of both signs, extended states can also be spin-glass-like for finite $x$, with the sequence of phases discussed above; {\it{i.e.}} from higher temperature paraelectric to (i) ferroelectric for $1 \ge x > x_{g}$, (ii) relaxor pseudospin glass for $x_{g} > x > x_{p}$, and (iii) remaining paramagnetic at all $T$ for $x_{p}>x>0$, but already with non-equilibrium nanodomain precursors above the transition temperatures for all  $0<x<1$, the size of the precursor region reducing to zero as the pure limits are approached \cite{foot-Burns}.

Many workers in the field of displacive relaxors conceptualize in terms of polar nanodomains that form above the  temperature of the susceptibility peaks,  with cooperative interactions between them eventually leading to the relaxor region. This conceptualization can be given substance by defining nano-moments in terms of negative eigenvalue eigenfunctions of Eq.(\ref{Anderson_equiv}), introducing them into an expanded partition function by adding them as variables with delta functions ensuring their identification, and then integrating out the original variables to give a system of interacting nanodomains. Technical practice would, however, be more challenging.

\subsection{Heterovalent relaxors}

\begin{figure*}[htb]%
\subfloat[Frequency-dependent a.c. susceptibilities of PMN; (1) 0.4 kHz ~- ~(6) 4.5MHz; reproduced with permission 
from 
G.\,A.~Smolenskii, V.\,A.~Isupov, A.\,I.~Agranovskaya and S.\,N.~Popov,
 Sov.Phys.Solid State \textbf{2}, 2584 (1961)
\copyright 1961 { AIP Publishing LLC} (http://aip.org)
.]
{%
\includegraphics*[width=.460\textwidth,height=.325\textwidth]{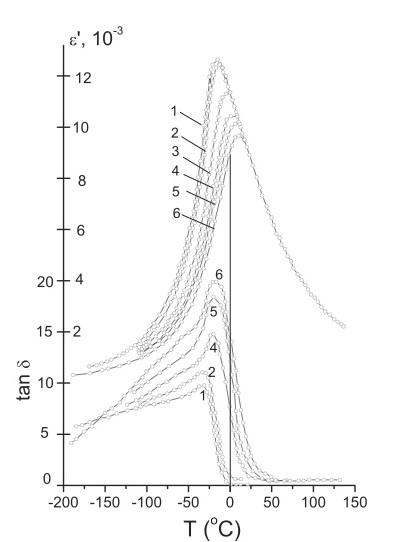}}\hfill
\subfloat[FC ($\Box$) and ZFC ($\circ$) static susceptibilities of PMN , together with remanent polarization on reheating ($\bullet$); 
reprinted with permission from 
 A.~Levstik, Z.~Kutnjak, C.~Filipi\v{c} and R.~Pirc, {Phys.Rev.B} {\textbf{57}}, 11204 (1998)
\copyright 1998 {American Physical Society} 
http://link.aps.org/abstract/PRB/v57/p11204
]{%
\includegraphics*[width=.460\textwidth,height=.3\textwidth]{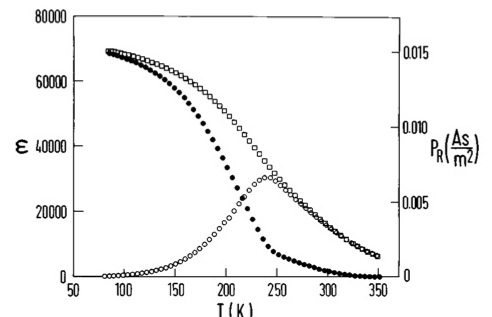}}\hfill
\caption{%
Susceptibilities of heterovalent relaxor alloy PMN.}
\label{fig:PMN_susc}
\end{figure*}

The situation for heterovalent relaxors in which the charge +4 $B$ ions of ${AB\rm{O}}_3$ are replaced by $B', B''$ of charges different from +4, but of overall average charge +4, is more complicated because they impose quasi-random excess charges compared with the corresponding template pure system, giving rise to random fields. 

However, the frequency-dependent susceptibility of PMN \cite{Smolenskii_3} (Fig $\ref{fig:PMN_susc}$a) is still very reminiscent of the case for spin glasses and there is still an onset of non-ergodicity, measurable through a separation of the FC and ZFC quasi-static susceptibilities \cite{Levstik} (Fig. $\ref{fig:PMN_susc}$b). It is therefore tempting to suspect that the origin is analogous to that proposed above for the isovalent relaxor BZT, carried by the $B$-site atoms, with the random fields and some of the displacements of the Pb and O atoms of a complementary origin \cite{Sherrington_PMN}.

The philosophy is to model the heterovalent system as having a Hamiltonian consisting of two parts, one  analogous to that considered above for the isovalent case with fictitious $B'^{*}, B''^{*}$ ions, most of whose properties are those of the real $B',B''$ ions with the exception of their charges, which are taken as 4+, together with a second term describing the difference between the real system and this effective isovalent system, in particular taking account of the charge differences of the real and fictitious ions and their influence on the system.

Thus, in the case of PMN, the fictitous effective isovalent system is $\mathrm {Pb(Mg}^*_{1-x}\mathrm{Nb}^*_{x}\mathrm{)O}_3$ (PM*N*) where Mg* and Nb* have the same properties as Mg and Nb, except for their charges. One could consider this fictitious system at any $x$, not just the $x=2/3$ of PMN. Noting that Mg++ and Zr++++ have essentially identical ionic radii (both 86 pm) and Nb+++++ and Ti++++ have similar ionic radii to one another (respectively 78 and 74.5 pm) and that the first of these pairs have ionic radii significantly larger than the second pair, it seems probable that PM*N* would behave analogously to BZT with Mg* replacing Zr and Nb* replacing Ti \cite{More_radii}. Thus one would expect that the Mg* ions would be effectively frozen at their normal locations but the Nb* ions able to displace in order to gain energy from cooperative ordering. This difference in ability to displace is in accord with common belief and confirmed in other alloys by first-principles calculation and simulation \cite{Bellaiche-Vanderbilt}. Thus, in PM*N* one would expect a range of equilibrium phases as a function of $x$ similar to that in BZT. Interestingly, the PMN value of $x=2/3$ is very close to the top of the relaxor range of $x$ in BZT \cite{Maiti} \cite{Shvartsman}.

The missing terms in the full PMN Hamiltonian are essentially given by the addition to the PM*N* Hamiltonian of terms due to charges of -2 at Mg sites and +1 at Nb sites. These will give rise to effective local fields, both on the $B$-sites and on the Pb- and the O- sites. Those on the $B$-sites represent an additional feature not found in spin glasses (since one cannot produce random fields in a magnetic system, except in the special guise of gauge-transformed uniform fields in an anti-ferromagnet). We return to these later. The fields on the Pb and O sites due to random $B$-site occupation by Mg or Nb will give rise to perturbing effects on those ions and, in the present author's view, are probably the origin of the quasi-spherical shell of distorsions of Pb that have been observed \cite{Vakhrushev}. Note, however, that to a first approximation these fields are independent of the relaxor distortions on the $B$-sites and so should not correlate strongly with the relaxor features.

\section{Strain glasses}
In this section we shall argue for another analogue of spin glasses,  of  strain distortions in martensitic alloys \cite{Sherrington_martensite} \cite{Sherrington_Glassy}. 

Martensitic materials, see e.g. \cite{Bhattacharya} 
\cite{Otsuka}, exhibit first-order structural phase transitions from higher
temperature phases of higher symmetry to lower temperature phases of lower
symmetry. 
One such example, on which we shall
concentrate initially for illustration, is from high temperature cubic austenite to a lower
temperature phase of alternating twin planes of complementary tetragonal character, epitomized by TiNi which in its pure state is a compound of rocksalt structure \cite{foot-note-name}. Our interest here will be particularly in when this compound is macroscopically atomically disordered, for example by altering the balance of Ti and Ni or by replacing some of these atoms by Fe.

Following a common practice, we shall employ a phenomenological Ginzburg-Landau modelling in terms of deviatoric strains compared with the higher-temperature higher-symmetry phase.  In a three-dimensional cubic system there are three mutually orthogonal tetragonal local distortion orientations. However, for conceptual simplicity we shall initially discuss a technically simpler two-dimensional analogue in which the higher temperature phase  is square and the lower temperature phase allows two orthogonal rectangular distorsions. 
The distorsions may conveniently be characterized quasi-locally by the deviatoric strain $\phi(\textbf{r})=[\epsilon^{11}(\textbf{r}) -\epsilon^{22}(\textbf{r})]/{\sqrt{2}}$ where $\epsilon^{\alpha\beta}(\textbf{r})$ is a component of the Lagrangian strain tensor \cite{Rasmussen}. For conceptual introduction, initially we shall ignore disorder-induced random local strain fields. The free energy will then have two main terms, one local and one non-local, of the form
\begin {equation}
F_{M}=F_{L} + F_{NL}
\label{eq:Martensite0}
\end{equation}
with 
\begin{eqnarray}
F_{L}= \int d{\textit{\textbf{x}}}
\{    
 A({\textit{\textbf{x}}},T)   
  {   \phi({\textit{\textbf{x}}})    ^2}
+ B({\textit{\textbf{x}}},T) 
| \nabla{\phi({\textit{\textbf{x}}})}|^2  \nonumber \\
 - C({\textit{\textbf{x}}},T) {\phi({\textit{\textbf{x}}}) ^4}
+ D({\textit{\textbf{x}}},T) {    \phi({\textit{\textbf{x}}})    ^6}
 \}, 
\label{eq:FL}
\end{eqnarray}
where $B, C$, and $D$ are all positive and with temperature dependences that are not of qualitative consequence, in contrast to the $A$ which reduce with temperature in a relevant fashion, 
and 
\begin{equation}
F_{NL} =- \int 
d{\textit{\textbf{x}}} d{\textit{\textbf{y}}} 
V({\textit{\textbf{x}}-{\textbf{y}}}) 
\phi({\textit{\textbf{x}}}) \phi({\textit{\textbf{y}}}),
\label{eq:FNL},
\end{equation}
%
which has its origin in the imposition of compatibility constraints on the strains (St.Venant's law), with  
%
\begin{equation}
V({\textit{\textbf{R}}}) \propto -\cos{(4 \theta{(\textit{\textbf{R}}})} / |{\textit{\textbf{R}}}|^2
\label{VStV}
\end{equation}
where $\theta{(\textit{\textbf{R}}})$ is the angle subtended by $\textit{\textbf{R}}$ with repect to the Cartesian coordinate system of the pure compound \cite{Lookman}.

We recognise $F_{M}$ as having the form  of $H_{SSG}$ (Eq. ({$\ref{eq:Hsg}$)) with parameters such as to give first order phase transitions as the $\{A\}$ are reduced with reducing temperature. We also note that the interaction $V({\textit{\textbf{R}}})$ is frustrated due to its antiferroelastic character in directions for which $\cos{(4 \theta{(\textit{\textbf{R}}}))}$ is positive. It can be discretized (in both displacement and location spaces) as a spin 1 Ising model, analogous to the Ghatak-Sherrington model $H_{GS}$  discussed in Eq. ($\ref{eq:GS}$), as
\begin{equation}
F_{MG}=\sum_{i}{\cal{D}}_{i} {S_{i}^2 }
-\sum_{ij} {\cal{W}}_{ij}
 S_{i}S_{j}; ~~S_{i}=0,\pm{1},
\label{eq:MG}
\end{equation}
where 
\begin{equation}
{\cal{W}}_{ij}=
V(\textit{\textbf{R}}_{ij}) +V_{SR}(\textit{\textbf{R}}_{ij})
\end{equation}
and $V_{SR}$ is a short-ranged ferroelastic interaction. In this formalism $S=0$ corresponds to the square austenite, $S=\pm 1$ to the two rectangular martensitic variants. For compactness we shall couch further discussion  in terms of the discretized model of Eq. ($\ref{eq:MG}$).


For the pure system, in which the  ${\cal{D(T)}}$ are uniform (site-independent, not disordered), decreasing as the temperature is decreased, minimization of $F_{M}$
 yields a first-order phase transition at a critical $T_c$ from the square (austenite) phase  for $T>T_c$ to a phase  of twinned ordering of  ferroelastic stripes of alternating rectangular distortion at angles of $\pi/4$ or $3\pi/4$, beneath $T_c$. This twinned (martensitic) phase is reminiscent of the striped phase of alternating $\pm$ spins which has been proven to be the low temperature phase  of an Ising model with a short-ranged ferroelastic
interaction and a long-ranged antiferroelastic interaction decaying with distance as a power law $(R/R_0)^{-p}$ with $d<p\le (d-1)$ \cite{Lebowitz} \cite{foot-Lebowitz}. The critical temperature occurs at a positive ${\cal{D}}$ at which the energy cost of the local deviation from $S=0$ is balanced by the energy gain from the site interactions with the martensitic $S=\pm1$.

Here we are particularly  interested in alloys, in which 
the ${\cal{D}}_{i}(T)$ vary from `site' to `site', dependent upon the local atomic environment, as well as depending upon temperature.  The minimum free energy state is given by \{$S_{i}=0,\pm{1}$\} depending upon whether  
$\{{\cal{D}}_{i} {S_{i}^2 }
-\sum_{j} {\cal{W}}_{ij}S_{i}
S_{j} \lessgtr 0 \}$ 
when the choices are made self-consistently for all sites, `greater than' yielding $S_{i}=0$, `less than' yielding $S_{i}=\pm 1$ with the choices of sign such as to minimise the global free energy.


We separate the effects of  composition and temperature through distributions whose shape characterises the spread due to composition  and whose mean depends on temperature, reducing as the temperature is reduced. 

For simplicity we consider first the case where  the distribution is two-valued:
\begin{equation}
P({\cal{D}})=x {\delta{{\cal{(D-D}}^{(1)}(T))}} +(1-x){\delta{({\cal{D-D}}^{(2)}(T)}}
\label{eq:binary}
\end{equation}
and assume that ${\cal{D}}^{(2)}(T)$ is always too large to permit bootstrapping of $S\neq0$ at a ${\cal{D}}^{(2)}$ site at any temperature. Thus, in the minimum free energy state,  the only deviations from $S=0$ will occur at ${\cal{D}}^{(1)}$ sites. The pseudospins \{$S_{i}$\} at these sites will thus be given by 
the minima of 
\begin{equation}
F_{MA}=\sum_{i\in{\{1}\}}{\cal{D}}^{(1)}_{i}(T) {S_{i}^2 }
-\sum_{ij \in{\{1\}}} {\cal{W}}_{ij}
 S_{i}S_{j}; ~~S_{i}=0,\pm{1},
\label{eq:MA}
\end{equation}
where the summations are restricted to sites with ${\cal{D = D}}^{(1)}$. 
Thus either \{$S_{i}=0$\} or \{$S_{i}=\sigma_{i}$\} where the \{$\sigma_{i}$\} are given by the minima of 
\begin{equation}
F_{MAI}=
-\sum_{ij \in{\{1\}}} {\cal{W}}_{ij}
 \sigma_{i}\sigma_{j}; ~~\sigma_{i}=\pm{1}.
\label{eq:MAI}
\end{equation}
A phase transition occurs when the the temperature is reduced beyond that for which the \{$S=0$\} austenitic phase has the lowest $F_{MA}$ to that with \{$S_{i}=\sigma_{i}$\} on the ${\cal{D}}_{1}$ sites, with the relative \{$\sigma_{i}$\} determined by the minimum of $F_{MAI}$.

\begin{figure*}[htb]%
\subfloat[ Schematic phase diagram for strain glass alloy with binary $\cal{D}$ distribution of Eq. ($\ref{eq:binary}$)]{
\includegraphics*[width=.425\textwidth,height=4.1cm]{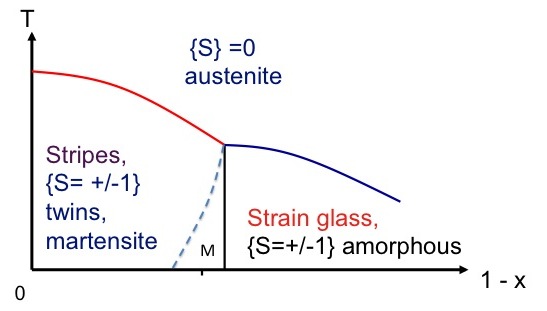}}\hfill
\subfloat[Experimental phase diagram of TiNi martensitic alloy ; from 
X-B. Ren et al., Phil.Mag. {\textbf{90}}, 141 (2010)
\copyright {Taylor and Francis} (http://www.informaworld.com) ]{%
\includegraphics*[width=.425\textwidth,height=4.1cm]{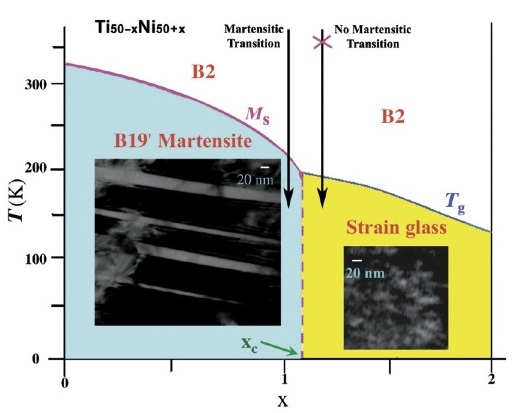}}\hspace{0.5cm}

\subfloat[Schematic phase diagram for strain glass alloy with Gaussian $\cal{D}$ distribution of Eq. ($\ref{eq:Gaussian}$).]{\label{figs5c}%
\includegraphics*[width=.430\textwidth,height=4.1cm]{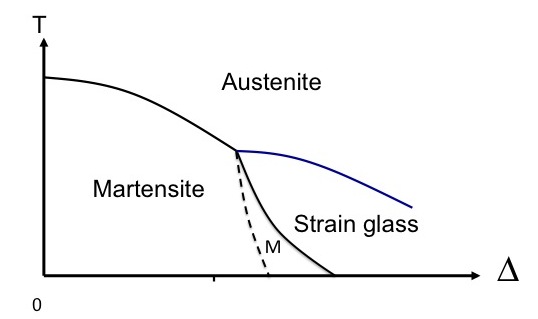}}\hfill
\subfloat[Phase diagram of TiNiFe martensitic alloy; 
reprinted with permission from 
J.~ Zhang, Y.~ Wang, X.~ Ding, Z.~ Zhang, Y.~Zhou, X-B Ren, K.~ Otsuka, J.~Sun and M.~ Song
 Phys.Rev.B {\textbf{84}}, 214201 (2011),
  \copyright {2011 American Physical Society}.
http://link.aps.org/abstract/PRB/v84/214201
 ]
{\label{figs4c5}%
\includegraphics*[width=.430\textwidth,height=4.1cm]{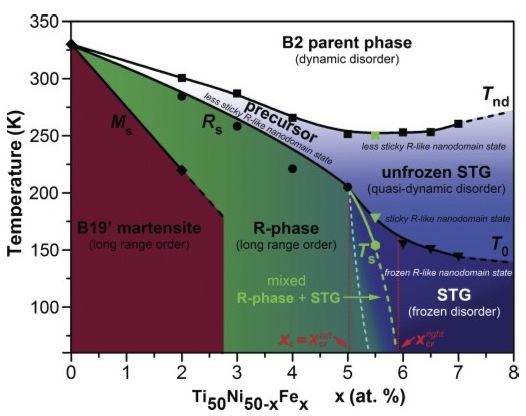}}\hspace{0.5cm}
\caption{%
Strain glass phase diagrams: (a) and (c) schematic predictions; (b) and (c) experimental results.
}
\label{fig:Strain_glass_phase_diagrams}
\end{figure*}

$F_{MAI}$ may now be recognised as an Ising example of $H_{CSG}$ (Eq.({\ref{eq:Hcsg}})) with $J_{ij} = {\cal{W}}_{ij}$. Hence the available bootstrapping binding energy corresponds to the gound state energy of this Ising system, itself estimable from its transition temperature. Thus, by comparison with known spin glass systems, it follows that the martensitic alloy's ordered phase will change from a twinned martensitic phase \cite{foot-martensite} to a  `strain glass' phase at a critical value of 
$x$, 
martensitic for $x>x_{c}$, 
strain glass for $x<x_{c}$ \cite{foot-mixed}; see Fig. $\ref{fig:Strain_glass_phase_diagrams}$(a). A similar behaviour has been observed in a ${\rm{Ti}}_{(50-x)}{\rm{Ni}}_{(50+x)}$ alloy ; see Fig. $\ref{fig:Strain_glass_phase_diagrams}$(b) \cite{Ren_2}, where the strain glass onset was signalled by the onset of a separation between FC and ZFC susceptibilities; see Fig $\ref{fig:FC_ZFC}$(a)  \cite{Ren}. 

Next we consider a situation where there can be a continuous range of \{${\cal{D}}$\}.
\begin{equation}
P({\cal{D}})={{((2{\pi})}\Delta^2)}^{-1/2}\exp{\{-[{\cal{D-D}}^{(0)}T)]^{2}/2{{\Delta}^2}\}}.
\label{eq:Gaussian}
\end{equation}
with ${\cal{D}}^{(0)}(T)$ reducing with reducing temperature. 
Clearly, there will be a phase transition when the temperature is reduced to a critical value sufficient to bootstrap \{$S\neq 0$\} on a macroscopic number of sites. 
Again, the lower temperature state will be either twinned martensite or strain glass depending upon the concentration $x$. However, in this case, as $T$ is reduced  beyond the initial onset of an ordered phase, more and more sites enter the range in which $S_{i} \neq 0$, increasing the effective concentration of magnetic sites in the pseudo-Ising spin glass of Eq. ($\ref{eq:MA}$). Hence, in this simple model, but similarly for other continuous ${\cal{D}}$ distributions, the phase line separating twinned martensite and strain glass should be re-entrant, moving towards  stronger disorder (larger $\Delta$) as the temperature is reduced. This predicted behaviour is presented schematically in Fig $\ref{fig:Strain_glass_phase_diagrams}$(c), 
along with the measured phase diagram for a TiNiFe alloy \cite{Zhang_et_al} in Fig $\ref{fig:Strain_glass_phase_diagrams}$(d) showing the predicted re-entrance, as well as a mixed phase ({\it{c.f.}} Fig. $\ref{fig:phase_diagrams}$(c)) \cite{foot-re-entrance}.

Clearly, within the strain glass phase many of the characteristic features of spin glasses can be anticipated with the simple `dictionary' transfers,  stress $\equiv$ external field, strain $\equiv$ magnetization.

\begin{figure*}[htb]%
\subfloat[FC and ZFC susceptibilities for strain glass; 
reprinted with permission from 
Y.~Wang, X-B.~Ren, K.~Otsuka and A.~Saxena, Phys.
Rev. B {\textbf{76}}, 132201 (2007); 
%
\copyright2007  {American Physical Society}
]{
\includegraphics*[width=.45\textwidth,height=4.5cm]{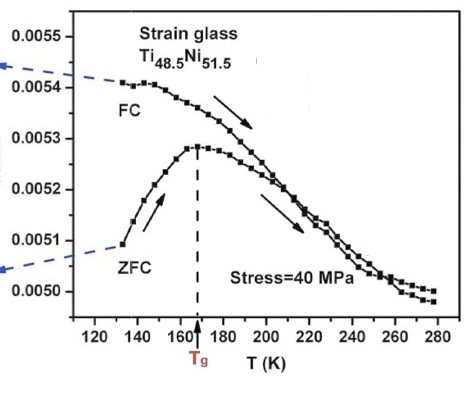}}\hfill
\subfloat[Schematic FC and ZFC susceptibitilties for a vector spin glass in a finite applied field; from 
D.~Sherrington, in {\it{Heidelberg Colloquium on Spin Glasses}}, eds. I.~Morgenstern and L.~van Hemmen, 125-136 (1983)
\copyright{Springer-Verlag} (www.springer.com). ]{%
\includegraphics*[width=.43\textwidth,height=4.7cm]{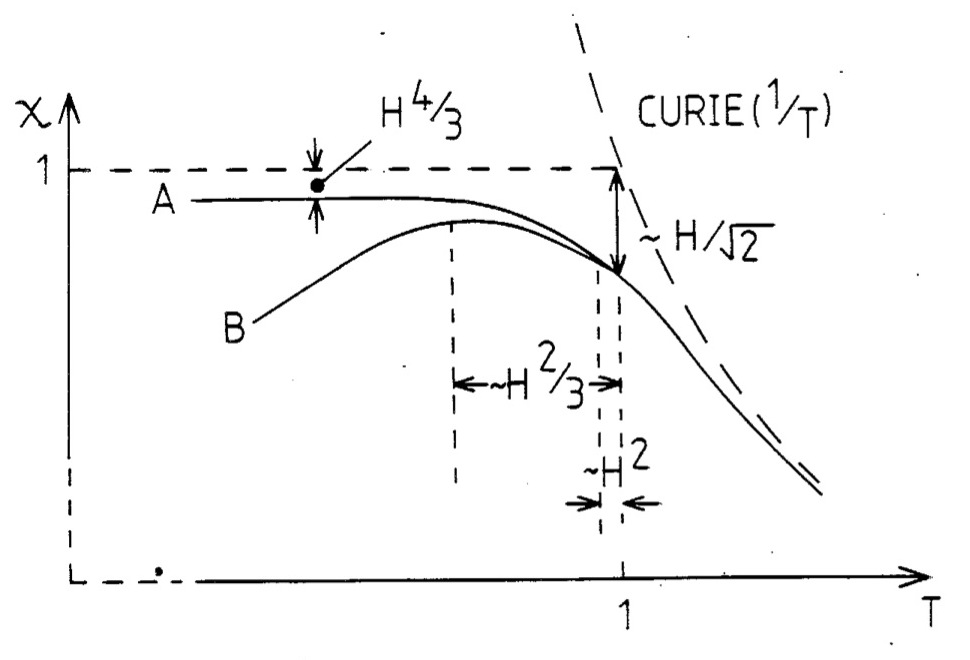}}\hfill
\subfloat[$M_{ZFC}$ and $M_{FC}$ for ${\mathrm{La}}_{0.5}{\mathrm{Sr}}_{0.5}
{\mathrm{CoO}}_{3}$ measured in finite fields; 
reprinted with permission from \cite{Nordblad2}  
D.\,N.\,M.~Nam, K.~Jonason, P.~Nordblad, N.\,V.~Khiem and N.\,X.~Phuc, Phys.Rev.B {\textbf{59}}, 4189 (1999)
\copyright 1999 {American Physical Society}
http://link.aps.org/abstract/PRB/v59/4189 (1999)
]{\label{figs4c6}%
\includegraphics*[width=.465\textwidth,height=4.65cm]{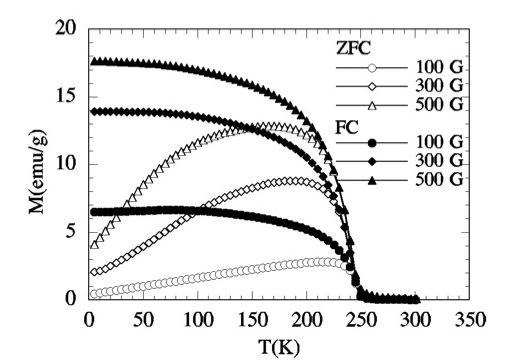}} \hfill
\subfloat[
Reduced temperatures (i) $\theta_{r}$ of separations of   $ M_{FC} and M_{ZFC}$ and (ii) $\theta_{a}$ of the peak in $M_{FC}$, as measured in figure (c).
%
]{\label{figs4c7}%
\includegraphics*[width=.450\textwidth,height=4.6cm]{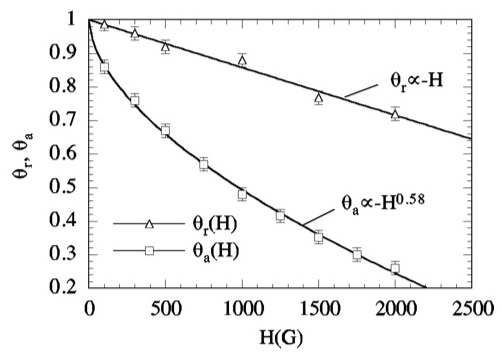}}\hspace{0.6cm}
\caption{%
FC and ZFC susceptibilities, (a) in a strain glass, (b) predicted for vector spin glass, (c) and (d) measured for a spin glass.}
\label{fig:FC_ZFC}
\end{figure*}

In the discussion above,  randomness in ${\cal{D}}$ corresponds to quenched disorder in the local atomic make-up of the alloys; {{\it{e.g.}} whether sites are occupied by the correct atoms of the pure compound or substituted by impurities in the alloy}, with the assumption that this only affects the tendency for a locally square structure or, equally probably, one of the two rectangular variants. If the alloying also introduces random strains, favouring one of the two rectangular variants over the other, then random field terms must be included too, as in 
\begin{equation}
F_{MGRF}=\sum_{i}({\cal{D}}_{i} {S_{i}^2 } -h_{I}S_{i})
-\sum_{ij} {\cal{W}}_{ij}
 S_{i}S_{j}; 
 ~~S_{i}=0,\pm{1},
\label{eq:MGRF}
\end{equation}
with some distribution ${\cal{P}}(h)$. Further discussion of this is, however, deferred until the next section.

So far, for  introductory simplicity, discussion was for a two-dimensional system. Extension to three dimensions is conceptually straightforward. In the continuous-$\phi$ language one needs only replace the scalar $\phi$ by a two-dimensional vector ${\boldsymbol{\phi}}$ that codes for the three orthogonal tetragonal martensitic orientations \cite{Rasmussen}. In the case of a system with $p$ martensitic orientations  a convenient representation is to use a generalized `Potts-{\it{cum}}-latticegas' model
\begin{equation}
F_{PLG}= \sum_{i}D_{i}(T)n_{i} 
-\sum_{ij} n_{i}n_{j}
\sum_{q=0}^{p-1}
{J^{q}
(\textit{\textbf{R}}_{ij})
\delta_{p_{i},p_{j}+q}}
\end{equation}
where $n_{i}=0,1$ is a lattice gas variable indicating whether a site is austenitic or martensitic and the $p_{i} =1, 2,..,p$ indicates the local martensitic variant. Potts spin glasses have further interesting features, at least within mean field theory; for a brief review see {\it{e.g.}} \cite{DS_Japan_2}). Notable among them
 is the prediction of re-entrance between spin glass and periodic phase as the temperature is reduced \cite{Elderfield_Potts} and of different types of complexity, dependent upon Potts dimension.

\section{Polar glasses}

There are also several analogues of spin glasses in diluted or otherwise quench-disorded polar or `orientational' systems with hard `quasi-spins', rather than the soft ones of the displacive systems discussed in section $\ref{section:Relaxors}$, both dipolar and quadrupolar. Examples are KCl:OH,  $\mathrm {(KBr)}_{1-x}\mathrm{(KCN)}_{x}$,
$\mathrm {(K}_{1-x}\mathrm{Li}_{x}\mathrm{)TaO}_3$, and ortho-para-hydrogen mixtures; for an experimental review see \cite{Hochli}.  The dipolar systems can be considered along essentially the same lines as the hard spin glasses above and exhibit similar features. Quadrupolar systems exhibit some new features compared with dipolar spin glasses, but similar to those of Potts glasses \cite{Goldbart}; for a theoretical review see \cite{Binder}, also \cite{DS_Japan_2}.        

\section{Fields; random internal and applied}

The discussion above has concentrated on limits of zero field. This is the case where the situation seems most clear. However, in both heterovalent relaxors and strain glasses effective random fields are surely present and in all cases strong uniform external fields can be applied. This topic raises many issues and they will not be explored thoroughly here. However, a few words seem appropriate. 

Consider first uniform fields. As noted at the beginning of this article, already in their important work in 1971 Cannella and Mydosh \cite{Mydosh} noted that even a very small field rounds out the peak in the susceptibility. However, it turns out that for the infinite-ranged Ising  spin glass model of Sherrington and Kirkpatrick \cite{SK} there is nevertheless a sharp (Almeida-Thouless( AT)) phase transition in a field \cite{AT}, characterised by  the onset of non-ergodicity \cite{MPV}; correspondingly, there is predicted a glassy phase within the ferromagnetic phase, called `Mixed' in Fig. $\ref{fig:phase_diagrams}$(c). 
 There has been much subsequent study and controversy as to whether this conclusion of a true, albeit unusual and subtle, phase transition carries over to finite-range systems. However, (i) the analogue systems above, both relaxors and strain glasses, have power-law-decaying effective interactions, far from short-ranged, and are frustrate), (ii) computer simulational studies on model one-dimensional spin glasses with random power-law distributed interactions have demonstrated that  crucial features
of the SK system extend also to power-law decaying systems with a small enough exponent in their denominators \cite{Sharma-Young2}, so it may be that a true sharp ergodic-nonergodic transition could exist in the relaxor and strain glass problems, and (iii) real experiments do not access the true thermodynamic limit and so could show effects even without a true transition. Hence, a corresponding glassy martensitic phase has been included in the predicted  schematic phase diagrams of Figs. $\ref{fig:Strain_glass_phase_diagrams}$(a) and   $\ref{fig:Strain_glass_phase_diagrams}$(c).


Also, in connection with measurements in finite uniaxial fields, it should be pointed out that for vector spins a different transition is expected to pre-empt strong longitudinal ergodicity-breaking. This is the Gabay-Toulouse (GT) phase transition \cite{Gabay-Toulouse} at which spin glass ordering onsets in the directions orthogonal to that of the applied field. Within (sophisticated spin-glass) mean field theory the prediction is that transverse spin glass ordering occurs at $T_{GT}(H)$, which decreases with $H$. The GT line is accompanied by the onset of strong tranverse ergodicity-breaking. Coupling terms   induce also  longitudinal ergodicity-breaking \cite{Cragg} but  this is initially much weaker than that transverse, becoming strong only below a crossover at a  lower temperature \cite{Cragg} \cite{GT_AT_scaling}. The anticipated consequences on FC and ZFC susceptibilities is illustrated (without explicit calculation) in Fig. $\ref{fig:FC_ZFC}$(b). It is tempting to wonder if this is, at least in part,  the origin of the similar feature of a gradual separation of still-rising FC and ZFC susceptibilities followed by a stronger separation only at a lower temperature, as seen in Fig. $\ref{fig:FC_ZFC}$(a) and, to a lesser degree, Fig. $\ref{fig:PMN_susc}$(b); i.e. to seek the explanation in a deduction that the fields applied in these measurements were not small enough for GT and AT to coincide (which strictly requires the limit of zero measuring field). In fact, behaviour reminiscent of that predicted in  \cite{Cragg}, \cite{Elderfield}, \cite{DS_Heidelberg} and  \cite{DS_Japan_2}, and illustrated in  Fig. $\ref{fig:FC_ZFC}$(b), was observed (almost two decades later) in \cite{Nordblad2} \cite{foot-Nordblad2}; see Figs.  $\ref{fig:FC_ZFC}$(c) and (d). These observations also make the author wonder whether some of the attributions of behaviour in several systems exhibiting FC/ZFC behaviour as in Fig. $\ref{fig:FC_ZFC}$(a) or Fig. $\ref{fig:FC_ZFC}$(c)  to `cluster glass' may in fact be manifestations of the effects of a large measuring fields in vector `spin' systems. 

Concerning random fields the situation is less clear. Many authors believe that  random fields are the drivers for the relaxor and/or the strain glass phenomena \cite{Kleemann} \cite{Cowley} \cite{Ren_5} but there is no good experimental magnetic example to compare with; the only examples are gauge-transformed dilute anti-ferromagnets in uniform fields which are very special in having only uniaxial effective fields and with randomness of sign correlated with the sublattices of the antiferromagnet \cite{Fishman}. Also, most analytic and simulational studies of random fields in magnetism have concerned themselves with systems where the exchange is ferromagnetic and mainly also short-ranged. There is, as yet,  no analytically soluble but non-trivial random-field model, analogous to that of the SK model for frustrated exchange disorder \cite{foot-nontrivial}. When there are random fields  in addition to SK-like random bond disorder  a sharp AT transition exists also for vector spins if the field randomness is isotropic in spin space \cite{Sharma-Young}, without a GT precursor \cite{foot-trivial}.

The similarity of the a.c. susceptibility peak structure in BZT and PMN is suggestive that random fields in the latter are not needed to drive relaxor behaviour \cite{foot-rf-BZT}. Alloys of PMN with   $\rm{ PbTiO_3}$ (PT),     $(PMN)_{1-y}(PT)_y$, are in accord with this suggestion since such alloying leads to an increase in the relaxor transition temperature, which corrolates with the facts that  (i) the concentation of active $B$-sites increases with $y$, because Ti and Nb are both active, and (ii) the concentration of extra charges, and hence random fields, decreases with $y$. 

As noted above, effective random fields are almost certainly also to be expected in martensitic alloys, reflecting anisotropic effects in the surrounding environments of any cell. This has been examined in a computer simulation study \cite{Ren_5} which includes only random-field and not random-anisotropy  (random $\cal{D}$) disorder and yet does demonstrate ergodicity-breaking effects similar to those expected for the random $\cal{D}$ case discussed above. Consequently, it is necessary to recognise that (i) the picture presented above needs further test and (ii)  it may be that either effective random-bond or effective random-field disorder may suffice (if strong enough) to produce  a glassy phase in suitable circumstances.

With regard to the possibility of strain glass consequence with only random-field disorder, we recall that, as noted earlier, most theoretical work on random field effects in magnetism has considered only systems that are unfrustrated without the random fields. In the case of the martensitic alloys this is not the correct scenario (as neither is it for the ferroelectric and relaxor systems discussed above.  Although not disordered, the bare interaction is frustrated and long-ranged. Just as does random dilution \cite{Fernandez}, the imposition of a random field prevents any simple uniform compromise in taking optimal advantage of exchange and field contributions to the energy and might, similarly to dilution, lead to the possibility of a spin glass-like compromise.

\section{Conclusions, predictions and tests}

This paper has been concerned with potential commonalties between ferroic glasses of several types, viewed through  the eyes of a spin glass theorist, employing similarities of skeletal models and conceptualization  in complementation of  comparisons of experimental observations. A particular emphasis has been given to transfers of spin glass understanding towards explanations and expectations for displacive relaxors in diluted frustrated ferroelectrics and the  strain glass phase of martensitic alloys. Some of these systems  have extra complicating ingredients that have not been included and which may play important roles with regard to some observables, or even be the actual drivers. However, it is hoped that this bare-bones approach will at least stimulate consideration of the ideas presented here and also lead to further tests and re-considerations of alternative cherished beliefs. 


Hence, I list here a few suggestions for further experiments and/or simulations to test some of the conceptualizations presented here.

(i) Investigate experimentally the low field limit FC and ZFC susceptibillities of BZT, across a range of relative concentrations, and  the possible correlation of an onset of a separation with the low frequency limit of the peak temperatures in the a.c. susceptibility measurements.

(ii) Investigate the effects on FC/ZFC separation of varying a finite  field on an isovalent relaxor such as BZT, either experimentally or simulationally. There could usefully be comparison with the schematic predictions of \cite{Cragg} and Fig $\ref{fig:FC_ZFC}(b)$. 

(iii) In comparison with spin glasses, investigate in BZT the remanent polarizations in the relaxor phase of BZT, comparing the isothermal (IRP) and thermoremanent (TRP) instances. 

(iv) Extend the simulations of Akbarzadeh et al.\cite{Akbarzadeh2012} on BZT  to include strong computer generated random fields to test their consequences and compare with PMN, also for a relative concentration closer to that of PMN.

(v) In complementation of the phase-field Ginzburg-Landau studies of \cite{Ren_5}, 
 investigate  within a Monte Carlo simulation whether adding random fields to  spatially frustrated exchange such as studied for dilution in \cite{Fernandez} would also have a spin glass consequence.

(vi) Perform analogues of the aging, re-juvenation and memory experiments of \cite{Vincent} on BZT, PMN and martensitic alloys within the relaxor or strain glass regimes, importantly noting that a temperature drop of several percent was needed in the spin glass case to exhibit the memory effects shown in Fig $\ref{fig:Vincent}$; such experiments have been reported on PMN \cite{Kircher} but with temperature steps of less than 1\%. 

Although we have argued for similar physics and mathematics, the transition temperatures (and correspondingly characteristic energies) for the onset of quasi-spin-glass behaviour are typically rather different for magnetic spin glasses and their structural analogues, as the figures presented above show, with those for the structural systems typically closer to room temperature. This could offer practical advantages in studying the structural systems to examine properties of the more general class of ferroic glasses. 

Also, the soft quasi-spin behaviour of displacive relaxors could provide a useful window to itinerant metallic spin glasses of the type epitomised by models such as those of Eqns. ($\ref{eq:Hubbard}$) or ($\ref{H_SM}$) with one constituent having too small a $U$ for itinerant ferromagnetism in the pure state  with a second that is an itinerant magnet in its pure state but has no local moment in isolation in a host of the first constituent.

It should be noted that the discussion presented above is not the first to use spin-glass type modelling to study either relaxors or martensitic alloys. For relaxors we may cite particularly the  work of Blinc and Pirc \cite{SRBRF} who proposed a spherical analogue of the SK model in a random field, motivated by a picture of nanodomains of several sizes interacting with both effective random bonds and random fields. For martensitic alloys we note the work of Krumhansl {\it{ et al.}} who also employed an SK-type model \cite{Krumhansl}, again effectively assuming the applicability of a random-bond spin glass model from the start.  Here, however, no such {\it{a priori}} assumptions have been made and the modelling retains the nuances of real (experimental) random-site frustrated exchange spin glasses. 

A  pseudo-spin approach to structural phase transitions was already used many years ago \cite{Stinchcombe},  but in a rather different fashion from that used here and not for highly disordered and frustrated systems with analogues of a spin glas state.

Relating PNR formation to that of localization in relaxors was also considered before \cite{Nambu} but for a system in which random bonds were assumed {\it{a priori}},  more than two decades after \cite{DS-Mihill}.

\section{Caveat}
The present author is not an expert on relaxors or martensitic alloys. Nor does he profess to know everything that has been done in spin glasses. He presents this document in the hope that it will be found stimulating, but he is conscious that there are in reality complicating factors, other possibilities and differences of opinion. He apologises to everyone of whose relevant work he is unaware and/or has failed to acknowledge.

\section{Acknowledgements}

The author thanks Avadh Saxena and Turab Lookman for introducing him to martensitic alloys and Xiaobing Ren for discussion and correspondence on that topic;  Roger Cowley for introducing him to relaxors and Wolfgang Kleemann, Rasa Pirc and Laurent Bellaiche for discussions and/or correspondence on that topic. He is also grateful to his many long-time colleagues and friends in the spin glass community, especially but not exclusively Scott Kirkpatrick, Giorgio Parisi, Marc M\'ezard, Nicolas Sourlas, Hidetoshi Nishimori, Ton Coolen, Michael Wong and Eric Vincent, for numerous helpful discussions. An award of an Emeritus Fellowship from the Leverhulme Trust is much appreciated.

\begin{figure}[htb]%
\hspace{1cm}
\subfloat{%
\includegraphics[width=.125\textwidth,height=.125\textwidth]{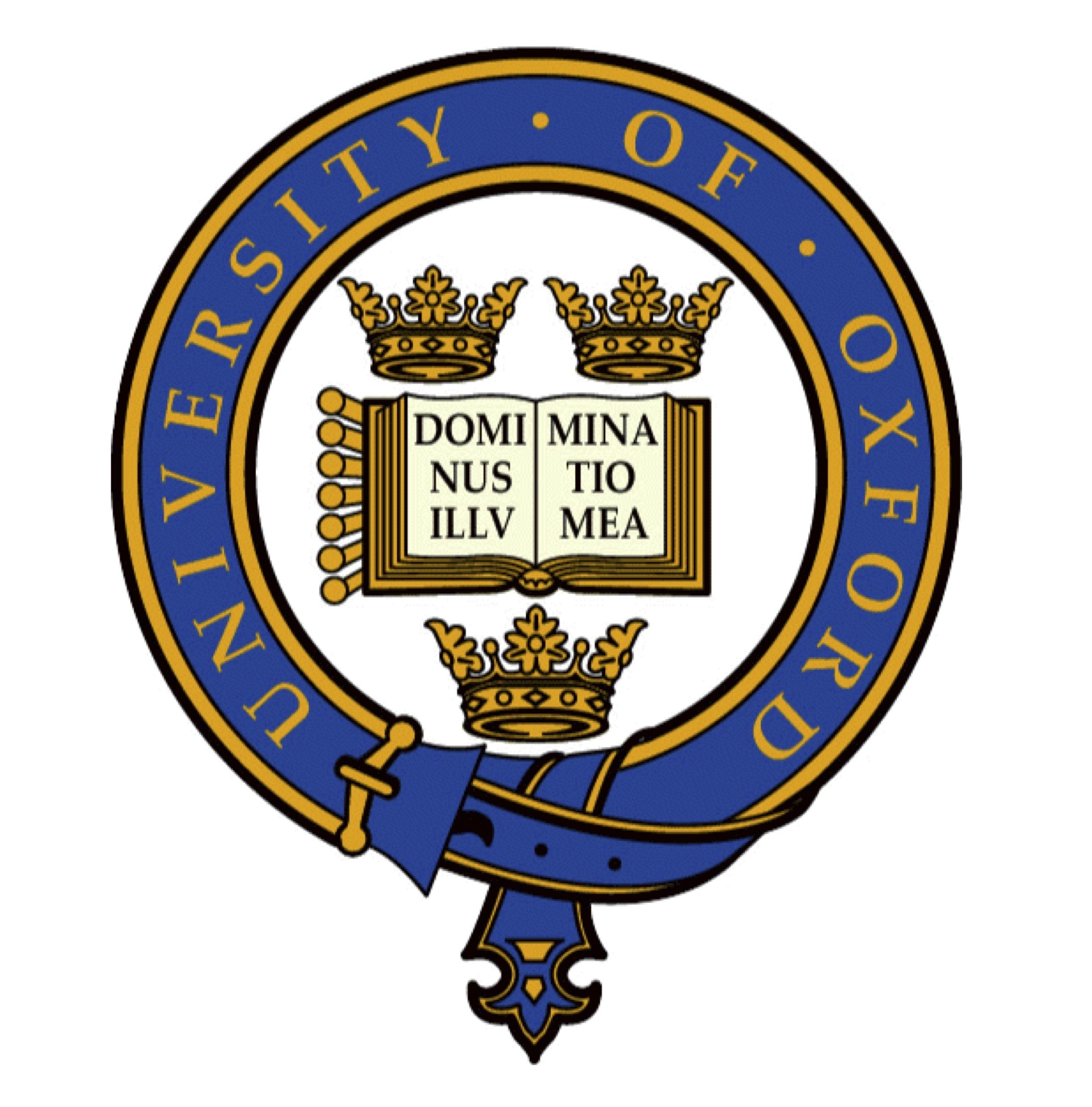}}\hfill
\subfloat{%
\includegraphics[width=.125\textwidth,height=.125\textwidth]{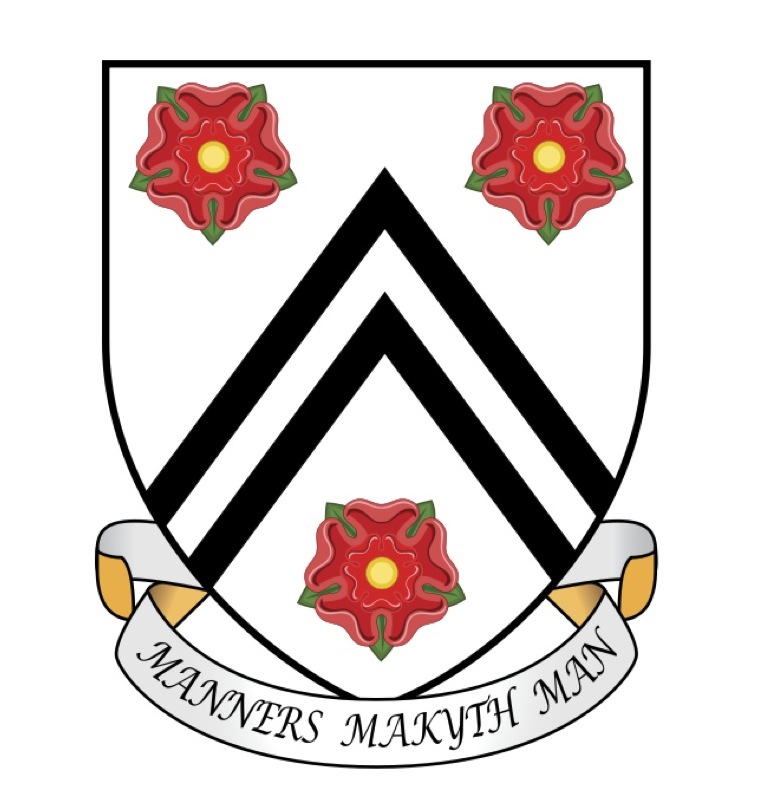}}\hfill
\subfloat{%
\includegraphics[width=.125\textwidth,height=.125\textwidth]{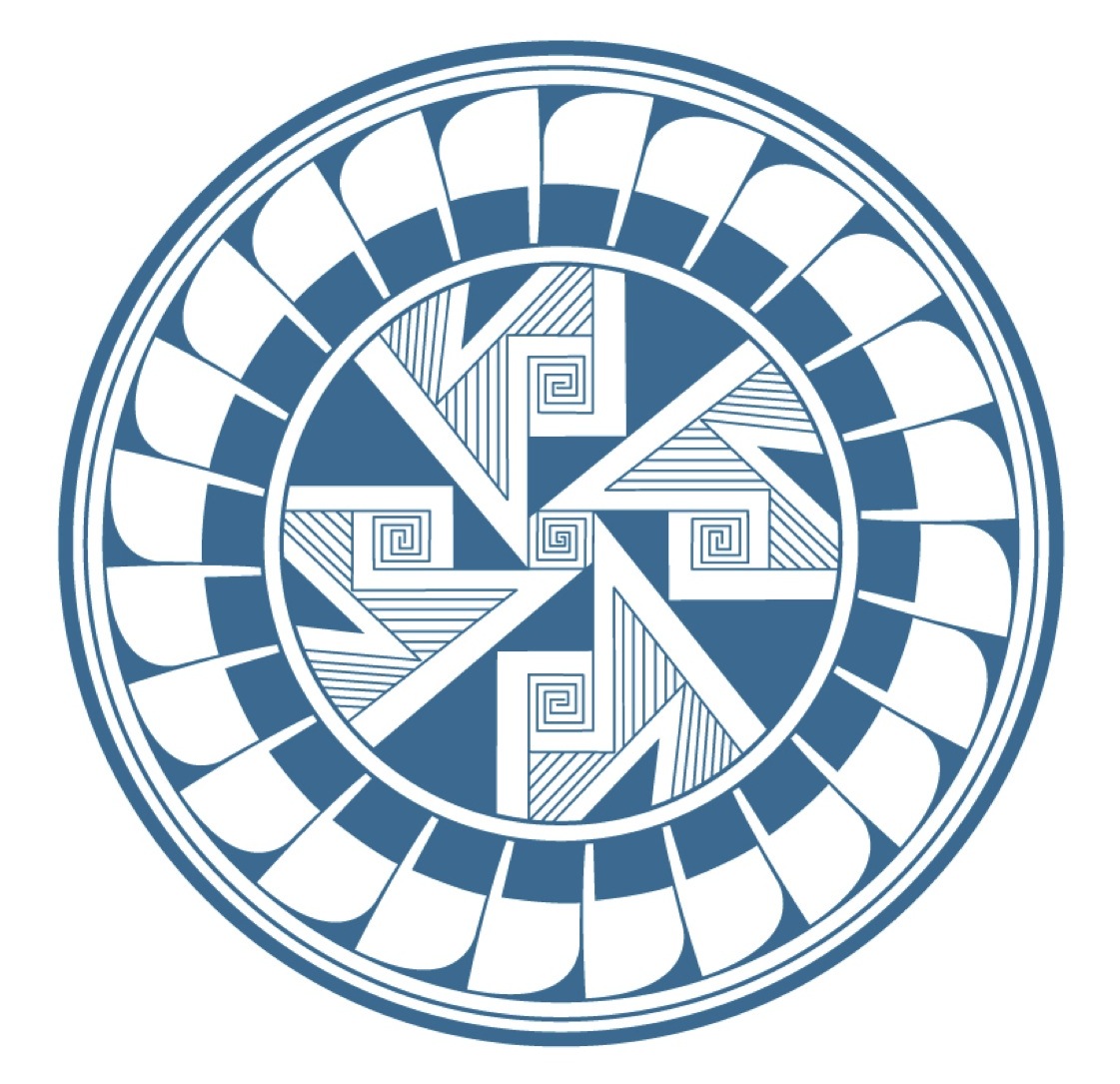}}
\hspace{.5cm}
\caption*{%
}
\label{fig:logos}
\end{figure}


\begin{thebibliography}{[1]}
\bibitem{footnote1} {In fact, frustration can  arise with purely anti-ferromagnetic interactions, for example via different neighbours, and even nearest-neighbour ferromagnetism can be frustrated if there are other constraints, such as the `spin ice rule'.}
\bibitem{foot-IRM-TRM}%
{For  the thermoremanent magnetization (TRM)
 the system is cooled in the field. For the  
  isothermal remanent magnetization (IRM),
 the field is applied only after cooling}
\bibitem{foot-strain}%
{Note that strain coupling is required to get the tetragonal crystal structure distortion of the ferromagnetic phase at larger $x$. However, the present interest is in the relaxor  phase which has neither overall polarization nor change in average crystal structure.}
\bibitem{foot:AF}%
{Note that an antiferromagnetic interaction of longer than nearest neighbour range can already be frustrated. As an example of spin glass behaviour with purely dipolar interaction in  combination with site-dilution, see \cite{Fernandez}.}
\bibitem{foot-strain2}%
{As noted earlier, the discussion above has excluded the strain coupling and absorbed the Ba and O ion effects into an effective $B-B$ interaction. To deal with the ferroelectric phase more completely these need to be included.}
\bibitem{More_radii}%
{The ionic radii of Pb++and Ba++ are respectively 133 and 149 pm. O-- is 126 pm. }
\bibitem{foot-note-name}%
{Note that even pure TiNi is often referred to as an alloy, as in `shape-memory alloy', but in fact it is a compound. When the expression `alloy' is used in this paper it refers to systems that deviate in a random fashion from a periodic structure.}
\bibitem{foot-Lebowitz}%
{Note, however, that actually the $V$ of Eq. ($\ref{VStV}$) (and its extension to higher dimension) decays with an inverse power law $p=d$ just beyond the realm of applicability of the proof of \cite{Lebowitz}, togethr with an extra angular factor. Hence boundary effects are more important in this pure case. But the main interest here is in the disordered alloy extension.}
\bibitem{foot-martensite}%
{Note that for $x\neq 1$ the twinned phase will have defects.}
\bibitem{foot-mixed}
{The SK model, exact for infinite-ranged interactions, has  a `mixed'  phase for an $x$-region above $x_c$, but such a phase in short-ranged spin glasses is contentious. In the martensitic alloys the interactions are power-law, which may suffice.}
\bibitem{foot-re-entrance}
Note that the phase lines in 
the strain glass predictions
 shown in the figures are schematic. In particular, the sign of the slope between the martensitic phase and the mixed phase in the figure for a continuous $\cal{D}$ is not calculated and hence its prediction is currently uncertain.
\bibitem{GT_AT_scaling}%
{For low fields the GT shift goes as $H^2$ and the AT crossover as $H^{2/3}$, where $H$ is the applied field.}
\bibitem{foot-Nordblad2}%
{This paper does not, however, cite these predictions.}
\bibitem{foot-nontrivial}%
{By `non-trivial' we refer to a situation where the random fields are essential for the phase transition. Thus, the (soluble) infinite-ranged ferromagnet with random fields is `trivial' in that the only phases are paramagnet and ferromagnet.}
\bibitem{foot-trivial}%
{Note, however, that the spin glass phase is not induced by the random fields, so this system is still `trivial' in the sense of \cite{foot-nontrivial} although the solution of this model requires subtle mathematics.}
\bibitem{foot-rf-BZT}%
{When Ti are displaced in relaxor BZT they do provide extra quasi-random fields but these are secondary to the interaction-driven terms discussed above.}
\end{thebibliography}
\end{document}